# Sizing and Dynamic modeling of a Power System for the MUN Explorer Autonomous Underwater Vehicle using a Fuel Cell and Batteries

**Mohamed M. Albarghot[1], Tariq Iqbal[2], Kevin Pope[3] and Luc Rolland[4]**

1, 3 Department of Mechanical Engineering, Memorial University of Newfoundland, St. John's, NL, Canada

2 Department of Electrical Engineering, Memorial University of Newfoundland, St. John's, NL, Canada

4 Department of Automation and controls, University of West Scotland, Scotland, UK

**Abstract**

The combination of a fuel cell and batteries has promising potential for powering autonomous vehicles. The MUN Explorer Autonomous Underwater Vehicle (AUV) is built to do mapping-type missions of seabeds as well as survey missions. These missions require a great deal of power to reach underwater depths (i.e. 3000 meters). The MUN Explorer uses 11 rechargeable Lithium-ion (Li-ion) batteries as the main power source with a total capacity of 14.6 kWh to 17.952 kWh, and the vehicle can run for 10 hours. The draw-backs of operating the existing power system of the MUN Explorer, which was done by the researcher at the Holyrood management facility, include mobilization costs, logistics and transport, and facility access, all of which should be taken into consideration. Recharging the batteries for at least 8 hours is also very challenging and time consuming. To overcome these challenges and run the MUN Explorer for a long time, it is essential to integrate a fuel cell into an existing power system (i.e. battery bank). The integration of the fuel cell not only will increase the system power, but it will also reduce the number of batteries needed as suggested by HOMER software. In this paper, an integrated fuel cell is designed to be added into the MUN Explorer AUV along with a battery bank system to increase its power system. The system sizing is performed using HOMER software. The results from HOMER software show that a 1-kW fuel cell and 8 Li-ion batteries can increase the power system capacity to 68 kWh. The dynamic model is then built in MATLAB / Simulink environment to provide a better understanding of the system behavior. The 1-kW fuel cell is connected to a DC / DC Boost Converter to increase the output voltage from 24 V to 48 V as required by the battery and DC motor. A hydrogen gas tank is also included in the model. The advantage of installing the hydrogen and oxygen tanks beside the batteries is that it helps the buoyancy force in underwater depths. The design of this system is based on MUN Explorer data sheets and system dynamic simulation results.

## I. Introduction

The MUN Explorer AUV is an autonomous underwater vehicle used for missions such as mapping, surveillance, oceanographic data gathering, environmental monitoring, mine detecting and coastal defenced [1]. One of the challenges facing the MUN Explorer is the power system's capacity to complete its missions. To improve the system's energy capacity, the MUN Explorer AUV is taken as a real example to do sizing and to build a dynamic model. The MUN AUV has a length of 5.3 m, a diameter of 0.69 m and a dry weight of 820 kg. In water, the flooded front and back sections of the AUV make the mass around 1400 Kg, with an average speed of 1.5 m/s, graphing over 80 Km. Some components have also been integrated into the vehicle such as computers and sensors.

Hydrogen production by Proton Exchange Membrane (PEM) water electrolysis is a promising method that has been successfully developed and integrated into renewable and hydrogen energy-based systems. Renewable energy sources, such as solar and wind, are desirable for hydrogen production due to random variations and significant current density capabilities [2]. PEM water electrolysis technology that generates hydrogen primarily emits water moisture, nitrogen and oxygen [3]. Energy storage or backup power systems are needed for photovoltaic and wind energy systems due to their discontinuous energy production. Batteries can be a good solution for daily storage but not for seasonal storage due to self-discharge. Storing energy in the form of hydrogen gas that is generated from renewable sources is a possible solution for both daily and seasonal storage [4]. For example, Sopian et al. (2009) integrated a Photovoltaic- wind- hydrogen energy production / storage system. The components of the system were a photovoltaic array, wind turbine, PEM electrolyzer, battery bank, and hydrogen tank. The system also had an automatic control system for battery charging and discharging. A hydrogen quantity of 130 ml/min to 140 ml/min was generated for an average global solar radiation between 200 $W/m^2$ and 800 $W/m^2$ and wind velocities ranging from 2.0 m/s to 5.0 m/s. For each system component, a mathematical model was built and compared to the experimental results [5]. Lithium–ion (Li-ion) battery technology has improved in the past decade. Li-ion batteries have higher energy and power density, higher efficiency and lower self-discharge when compared to other batteries (NiCd, NiMH, and Lead Acid). To ensure the Li-ion battery is operating at a proper temperature and state of charge (SOC), a battery management supervision system (BMSS) must be applied [6]. Fuel cells' high energy density, quiet operation, and high efficiency have allowed them to be used as a portable energy source. The capacity of fuel cells increased worldwide from 65 MW in 2009 to 181 MW in 2014 [7, 8]. Many types of fuel cells such as the proton exchange membrane fuel cell, alkaline fuel cell, and phosphoric acid fuel cell use hydrogen as fuel to produce electricity and water. Hydrogen-specific energy is high compared to other fuels' specific energy. Fuel cells have many applications such as stationary, transportation, and portable applications. Proton exchange membrane fuel cells have a higher efficiency compared to phosphoric acid fuel cells and alkaline fuel cells [9].

Using compressed hydrogen in composite cylinders for fuel cells is an alternative for underwater vehicles. Composite cylinders have a low weight and can increase the total performance of a deep-diving AUV. Furthermore, hydrogen cylinders may help buoyancy compensation in underwater depths. The design for underwater depths makes the weight of the pressure hull increase, and as a result, the amount of energy carried in a vehicle with neutral buoyancy is minimized with the design depth. Considering this, the batteries inside the vehicle should be as light as possible [10]. AUV energy supply powered by a fuel cell has been integrated on an IFREMER survey AUV called IDEFX by HELION, an AREVA Renewable subsidiary. Several experiments have demonstrated the interest in underwater power sources by installing a fuel cell along with a hydrogen gas tank [11].

This paper aims to design, size, and integrate a fuel cell into an existing power system that uses a battery bank as the main energy source to power the MUN Explorer AUV. By adding a fuel cell into the MUN Explorer, the power system capacity will be increased. However, the weight and the number of batteries can be reduced accordingly, and the number of hours of operation will increase. In this work, the focus will be on the main four components: the oxygen and hydrogen tanks, PEM fuel cell, Li-ion battery and DC motor (load). This paper is divided into three sections: the first section illustrates the components and system sizing using Hybrid Optimization Model for Electrical Renewable (HOMER) software; the second section demonstrates the dynamic modeling, simulation and results; and the third section is the conclusion.

## II. Components and System Sizing

### 1. Hydrogen / Oxygen Tanks and PEM Fuel Cell

The MUN Explorer Autonomous Underwater Vehicle as shown in Fig. 1 has plenty of vacant space that could be used to install the hydrogen and oxygen tanks as well as the fuel cell.

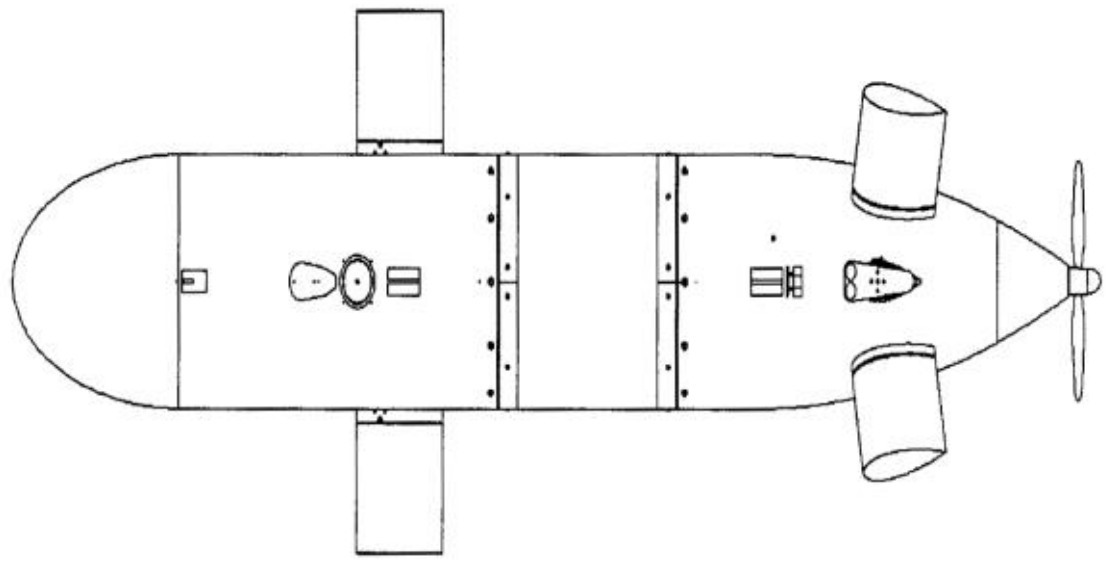

Fig. 1: Hull structure of the MUN Explorer AUV

The hydrogen consumed by the Proton Exchange Membrane Fuel Cell (PEMFC) can be generated directly from the electrolyzer. The hydrogen gas also depends on the relationship between the output power and the hydrogen needed for the PEMFC system. Excess hydrogen is directed to the storage tank. Due to the lack of an oxygen gas underwater surface, the fuel cell operation in underwater vehicles requires oxygen gas storage to complete the reaction between the cathode and the anode. By carrying the oxygen into the AUV, the fuel cell performance is increased by 2 to 3 times. To remove the produced water from the fuel cell during the operation, extra oxygen must be brought into the vehicle. This should be measured when the sizing of the oxygen storage is completed [12]. There are many ways to store hydrogen and oxygen. For example, compressed gas or liquid hydrogen and oxygen can be applied. HOMER software is designed to deal with renewable / non-renewable energy components and integrate them with each other. HOMER works by providing inputs (i.e. capital cost and size to consider kW) and design information about any given power system. HOMER simulation will give the system configurations and then create a list of feasible system designs and sort that list according to cost-effectiveness. Finally, a sensitivity analysis can be performed. The complete HOMER block diagram is illustrated in Fig. 2.

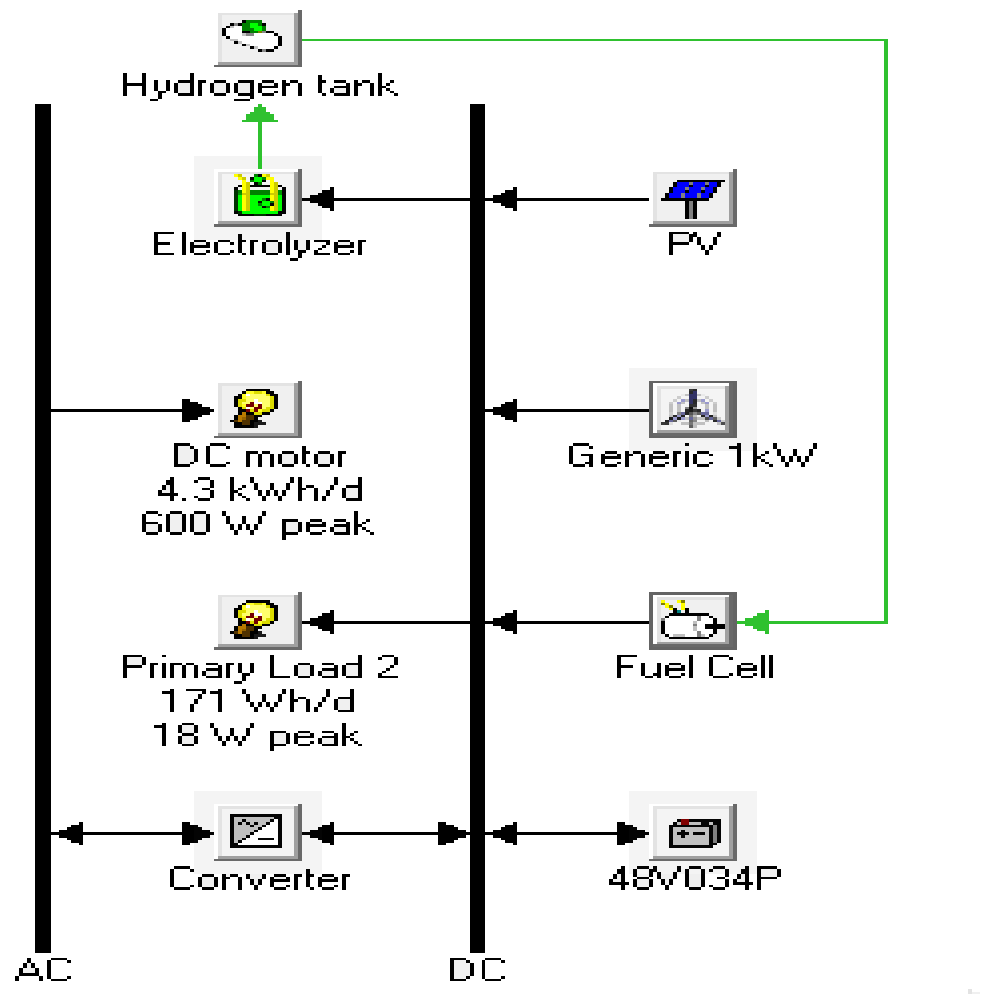

Fig: 2 HOMER block diagram

This diagram consists of renewable energy sources such as solar and wind to generate electricity to power the electrolyzer and then charge the battery. After that, the electrolyzer will generate the hydrogen and oxygen gases. Finally, the fuel cell and the battery will power the DC motor. It is understood that the wind energy, solar energy and electrolyzer will be onshore, and hydrogen and oxygen will be transferred to the AUV when it is docked. To run the HOMER software, the capital cost (i.e. commercial prices) of three different hydrogen tanks along with the sizes to consider (kg or kW) need to be entered into the hydrogen tank inputs. However, the reason for selecting three or more different hydrogen tanks is to give HOMER software more options to choose from so it can select the most optimal results. The same procedure is done for the fuel cell inputs, electrolyzer inputs, battery inputs, convertor inputs, PV inputs, and wind turbine inputs. The data sheet of each input and its price can be found in the attached appendix. Fig. 3 shows the simulation result of HOMER software in terms of the hydrogen tank storage level in (kg) and monthly statistics as well as frequency histogram. Since HOMER software does not have an oxygen tank input, the sizing will only be performed analytically in the next sections. After the capital cost and sizes to consider (0.2kW, 0.3kW and 0.5kW) have been set for the fuel cell inputs, the simulation runs to give the results as shown in Fig. 4. The values in the gray line have been chosen from HOMER software. The FC results are shown in Fig. 5.

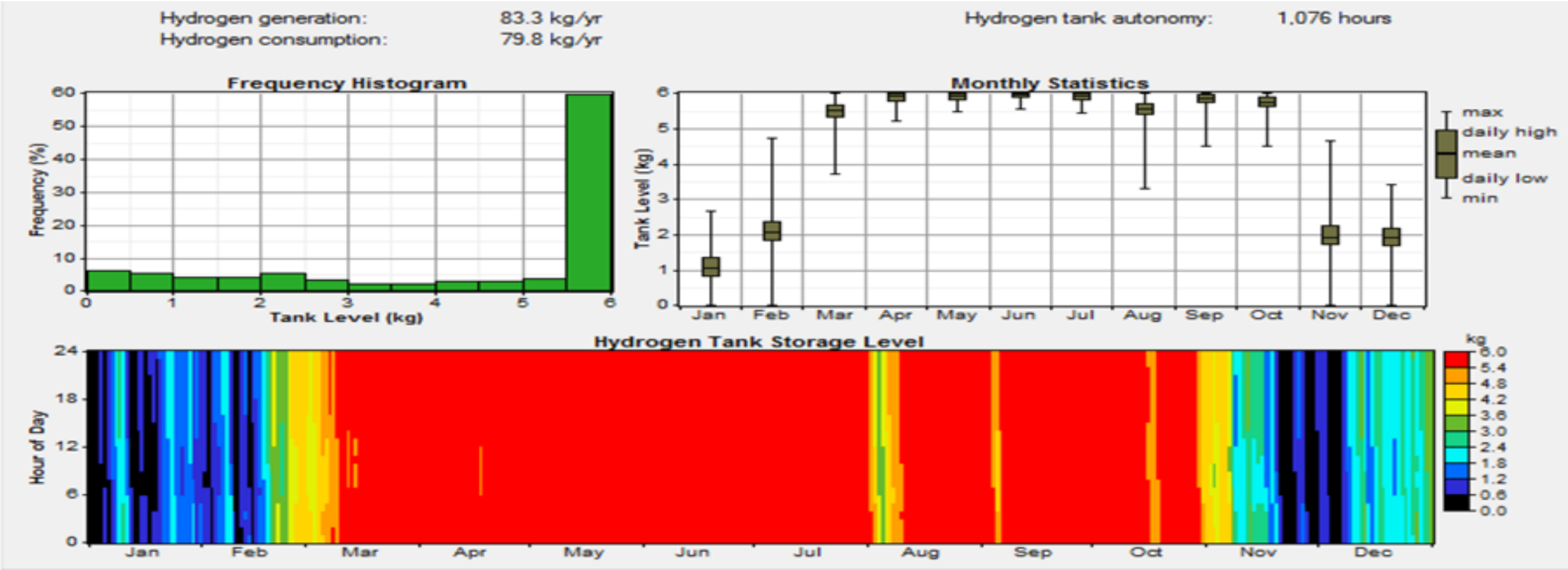

Fig. 3: Simulation results from HOMER software for the hydrogen tank

| | PV (kW) | G1 | FC (kW) | 48V034P | Conv. (kW) | Elec. (kW) | H2 Tank (kg) | Initial Capital | Operating Cost ($/yr) | Total NPC | COE ($/kWh) |
|---|---|---|---|---|---|---|---|---|---|---|---|
| | 7.02 | 1 | 1.0 | 10 | 2.0 | 2.4 | 6.000 | $ 109,276 | 760 | $ 118,997 | 5.717 |
| | 5.46 | 1 | 0.5 | 11 | 1.5 | 1.2 | 7.000 | $ 96,130 | 1,792 | $ 119,036 | 5.721 |
| | 7.02 | 1 | 0.5 | 11 | 2.5 | 1.2 | 6.000 | $ 98,769 | 1,589 | $ 119,079 | 5.721 |
| | 6.24 | 1 | 0.5 | 8 | 2.5 | 2.4 | 6.000 | $ 101,409 | 1,385 | $ 119,113 | 5.726 |
| | 7.80 | 1 | 0.5 | 9 | 1.5 | 1.2 | 6.000 | $ 100,089 | 1,490 | $ 119,140 | 5.725 |
| | 7.80 | 1 | 1.0 | 8 | 2.0 | 2.4 | 6.000 | $ 111,196 | 626 | $ 119,199 | 5.729 |
| | 5.46 | 1 | 0.5 | 11 | 1.0 | 2.4 | 6.000 | $ 99,429 | 1,548 | $ 119,217 | 5.728 |
| | 6.24 | 1 | 0.5 | 9 | 1.5 | 2.4 | 6.000 | $ 101,649 | 1,381 | $ 119,297 | 5.733 |
| | 7.02 | 1 | 1.0 | 11 | 1.5 | 2.4 | 6.000 | $ 109,816 | 744 | $ 119,327 | 5.733 |
| | 7.02 | 1 | 1.0 | 10 | 2.5 | 2.4 | 6.000 | $ 109,576 | 763 | $ 119,331 | 5.733 |
| | 5.46 | 1 | 0.5 | 11 | 2.0 | 1.2 | 7.000 | $ 96,430 | 1,795 | $ 119,370 | 5.737 |
| | 7.02 | 1 | 0.5 | 11 | 3.0 | 1.2 | 6.000 | $ 99,069 | 1,591 | $ 119,413 | 5.737 |

Fig. 4: Suggested results simulation by HOMER software

| Quantity | Value | Units |
|---|---|---|
| Hours of operation | 1,226 | hr/yr |
| Number of starts | 280 | starts/yr |
| Operational life | 122 | yr |
| Capacity factor | 3.20 | % |
| Fixed generation cost | 0.0270 | $/hr |
| Marginal generation cost | 0.00 | $/kWh |

| Quantity | Value | Units |
|---|---|---|
| Electrical production | 280 | kWh/yr |
| Mean electrical output | 0.228 | kW |
| Min. electrical output | 0.00126 | kW |
| Max. electrical output | 0.681 | kW |

| Quantity | Value | Units |
|---|---|---|
| Hydrogen consumption | 79.8 | kg/yr |
| Specific fuel consumption | 0.285 | kg/kWh |
| Fuel energy input | 2,661 | kWh/yr |
| Mean electrical efficiency | 10.5 | % |

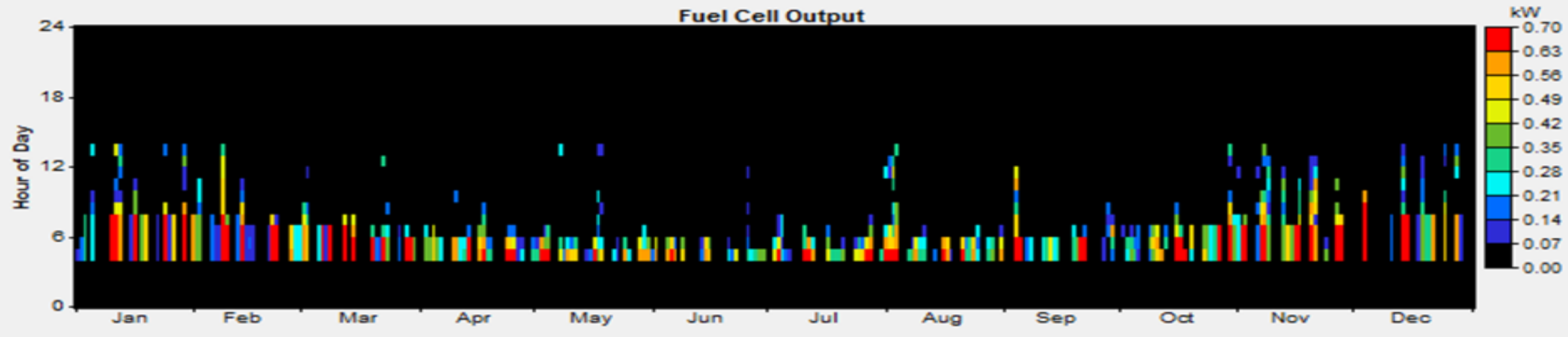

Fig. 5: HOMER software results for the fuel cell inputs

## 2. Lithium-Ion Battery and Converter

The MUN explorer uses Li-ion batteries as its main source of energy to power loads, which include all electronics onboard and the emergency lights. That is because these batteries have high energy density and efficiency compared to other types of batteries. A Li-ion battery is more attractive in portable applications such as automotive and autonomous vehicles.

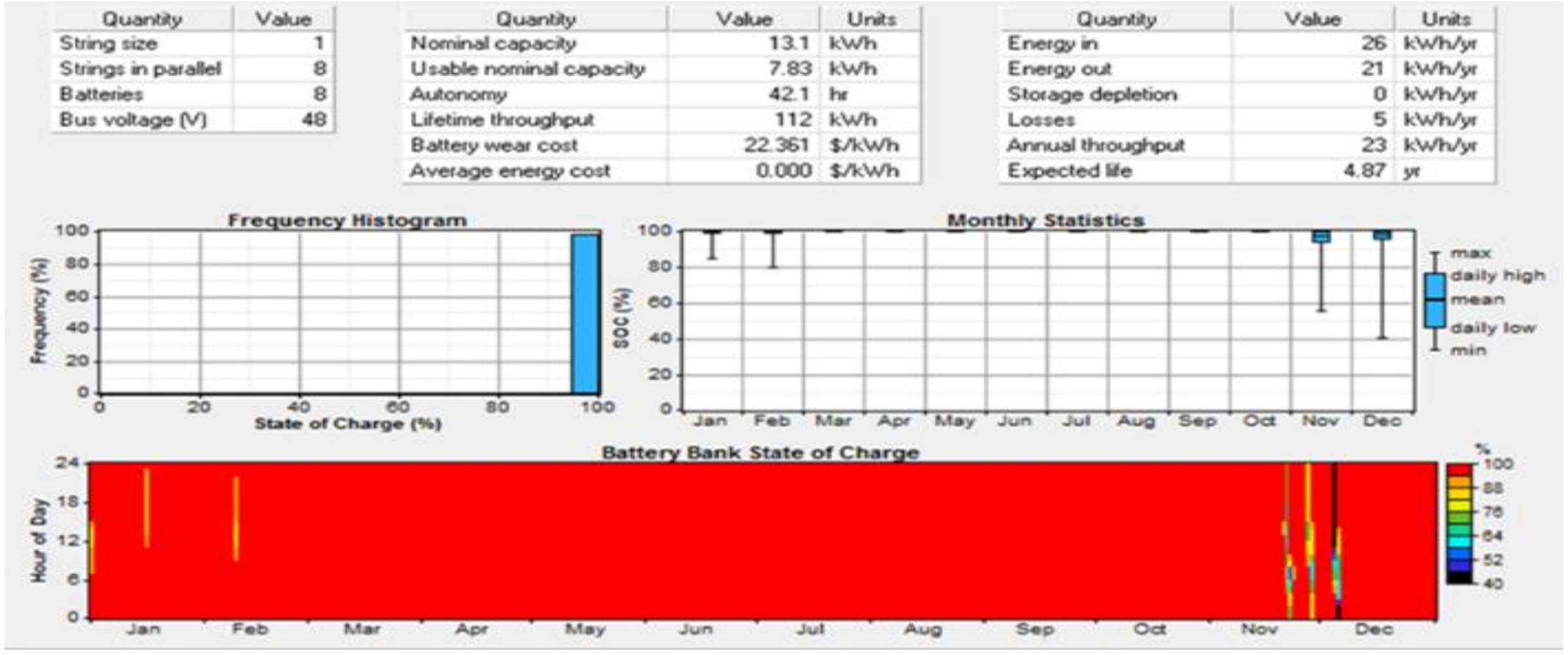

Fig. 6: HOMER software results for the battery

The cost of the Li-ion batteries and size to consider (i.e. number of batteries) have been entered into HOMER software inputs. Fig. 6 shows the battery characteristic results. The battery has a nominal voltage of 48 V and nominal capacity of 34 Ah. The DC bus of the system is set to be 48 V, which means the battery also must be 48 V. Those characteristics were provided by the battery's data sheet as well. The DC / DC boost converter is well known as a step-up converter, which takes a lower voltage to a higher voltage. The HOMER results suggested that a 2 kW DC / DC converter should be used in the system. The efficiency of a DC / DC converter is always above 90%, and it has a lifetime up to 15 years.

### 3. Permanent Magnetic DC Motor (PMDC)

In this case, the PMDC motor represents the load in HOMER software, and it is powered by the fuel cell and the battery. Permanent magnetic direct current (PMDC) motors are electrical machines that convert direct current electrical energy into mechanical energy. They are commonly used in many industrial, residential, and commercial applications [13]. The MUN Explorer AUV runs for ten (10) hours, so that the load has been specified based on the hours of operations (i.e. 10 hours) to be 600 W as illustrated in Fig. 7. The load is also divided into two sections: a DC load, which represents the electronics onboard, and the AC load, which is a variable speed motor. The MUN Explorer has only DC components, so the reason for selecting AC in HOMER is to represent the motor drive in our sizing.

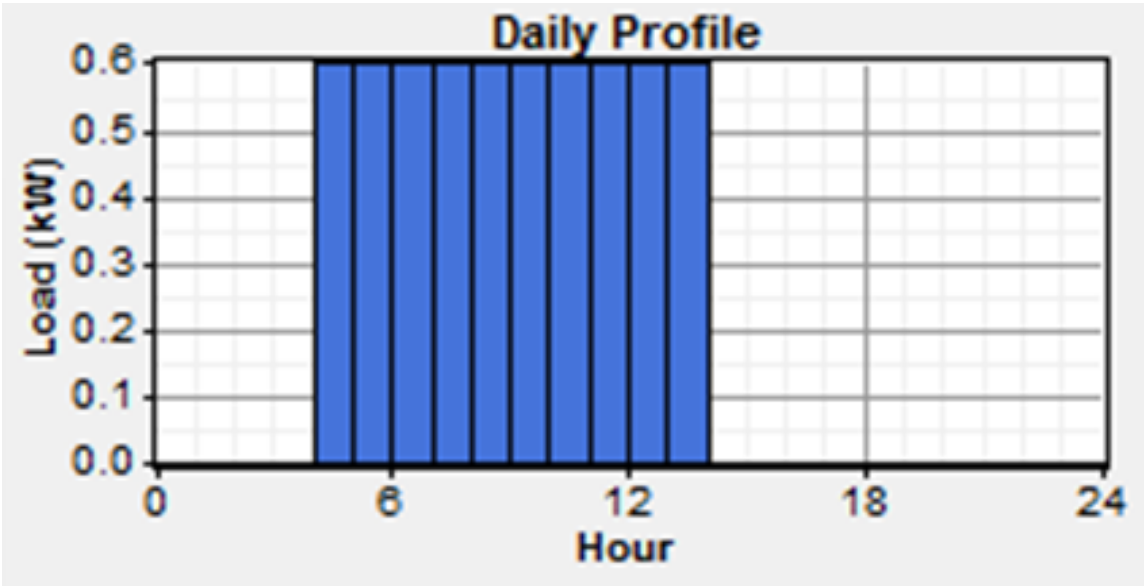

Fig. 7: The load for the DC motor in HOMER

## III. System Dynamic Model

### 1) Hydrogen / Oxygen Tank and PEM Fuel Cell

The storage system in the MUN Explorer could be challenging to install. As mentioned above, there are many ways to store compressed or liquid hydrogen and oxygen, especially for the MUN Explorer applications. Compressed or liquid hydrogen and oxygen storage gases can be implemented in terms of specific energies and energy densities. Effective storage systems that have higher energy density (ED) and specific energy (SE) are preferred [12].

The advantages of compressed hydrogen do not need preprocessing and are the easiest and cheapest solution for dealing with fuel storage. However, to maximize hydrogen content, high pressures (up to 700 bar) can be applied due to the low energy density of hydrogen gas. Liquid hydrogen has a higher density than gas. Liquid hydrogen also needs a temperature that is less than 20.15 K, so the stored liquid hydrogen must be in cryogenic Dewars (multi-shell flasks using an evacuated interstitial space) to eliminate heat transfer throughout the flask and prevent gas from reaching the boiling stage. Table 1 shows the specific energy and the energy density for compressed and liquid hydrogen storage systems, respectively [12].

Table 1: Hydrogen storage system for SE and ED

| Hydrogen | Specific Energy (kWh/kg) | Energy Density (kW/L) |
|---|---|---|
| Compressed | **1.71- 1.82** | **0.56 - 0.82** |
| Liquid | **2.05** | **1.86** |

Lightweight tanks for transporting the compressed oxygen applications are more accessible than the hydrogen ones because hydrogen tanks are used in automotive vehicle applications, while oxygen tanks are often used in medical applications. In short, hydrogen tanks can be modified for oxygen storage systems [14]. "Since high-pressure oxygen has a simple delivery mechanism, the desired oxygen tank wall thickness increases with pressure, which causes a reduction in the energy density advantages" [15]. Liquid oxygen storage can be a suitable solution for limited space applications. Some drawbacks of this storage system are its complexity due to the safety concerns associated with the handling and refueling process [15]. A liquid oxygen storage system prototype has been designed by Sierra Lobo, Inc. with a diameter of 54 cm (21 inches) [14]. This prototype can store 50 kg of liquid oxygen at 452 k to run a 1-kW output PEM fuel cell. The system is 0.94-m long and 0.32 m in diameter. The weight is 13.6 kg when it is empty and 63.6 kg when it is full [16]. Table 2 shows the specific energy and the energy density for compressed and liquid oxygen storage systems, respectively [14].

Table 2: Oxygen storage system for SE and ED

| Oxygen | Specific Energy (kWh/kg) | Energy Density (kW/L) |
|---|---|---|
| Compressed | **0.77 - 1.68** | **0.6 - 1.09** |
| Liquid | **2.9 - 3.3** | **2.78 - 2.98** |

In this paper, the model for the compressed oxygen / hydrogen tank corresponds to the one used by [17] and [18]. The dynamic model of oxygen / hydrogen tank was built based on equations (1) and (2) in the MATLAB / Simulink environment as shown in figures 8 and 9. The compressibility factor is defined as a function of temperature and pressure. Its value equals 1 when the pressure is less than 2000 psi and is higher than 1 when the pressure is higher than 2000 psi at room temperature [18].

$$P_b - P_{bi} = z * \frac{N_{H2} R T_b}{M_{H2} V_b} \quad (1)$$

$$Z = \frac{P V_m}{RT} \quad (2)$$

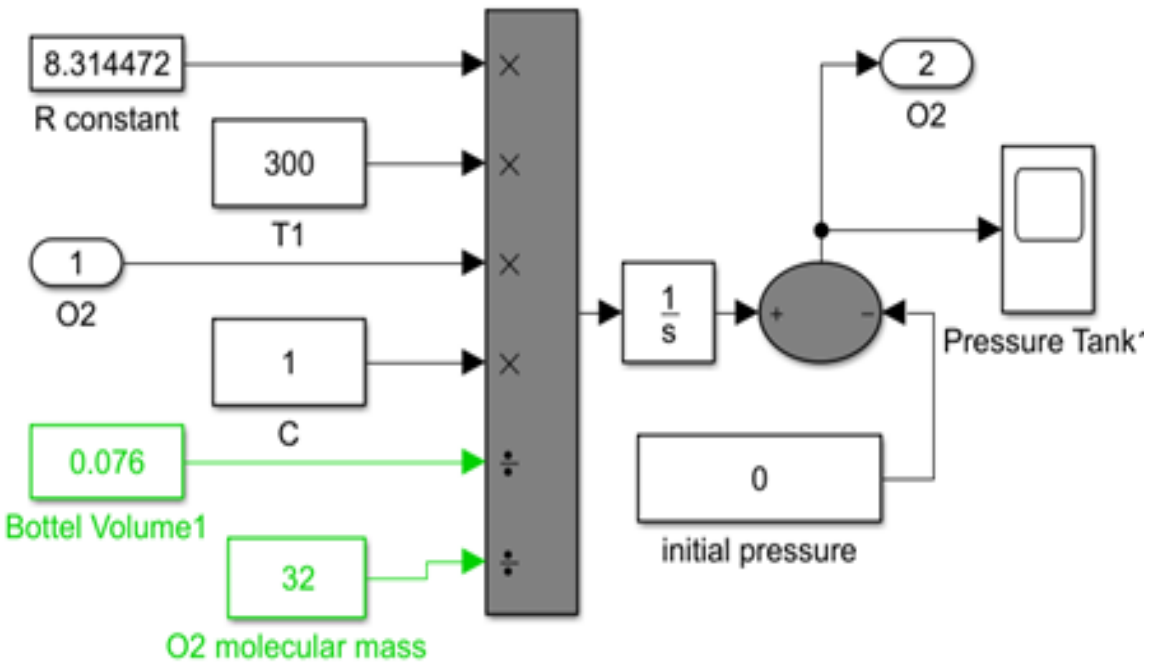

Fig. 8: Oxygen tank in MATLAB / Simulink [17 and 18]

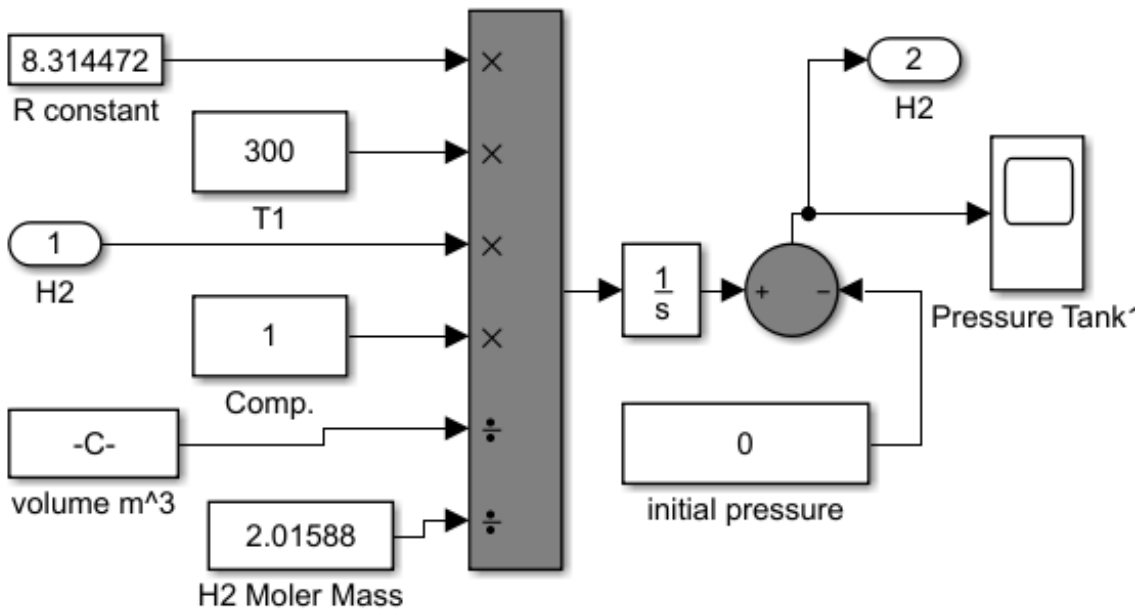

Fig. 9: Hydrogen tank in MATLAB / Simulink [17 and 18]

To evaluate the fuel cell in terms of specific energy and energy density, a commercial fuel cell (Horizon 100W PEM Fuel Cell) is integrated into the storage systems. It is selected due to the effectiveness of Horizon fuel cells and its recognised experience in AUV fuel cell applications. Table 3 shows the fuel cell parameters.

Table 3: Fuel cell parameters

| **Weight (kg)** | **Dimensions (cm)** | **Volume (L)** | **Specific Power (W/kg)** | **Power Density (W/L)** |
|---|---|---|---|---|
| **4** | **23.3*26.8* 12.3** | **7.68** | **250** | **130** |

The calculation of the ED and SE of the complete storage system is represented as follows [12]:

$$ED_{ss} = \frac{ED_{H2} * ED_{O2}}{ED_{H2} + ED_{O2}} \quad (3)$$

$$SE_{ss} = \frac{SE_{H2} * SE_{O2}}{SE_{H2} + SE_{O2}} \quad (4)$$

Those equations were applied for reactant storage combinations of liquid hydrogen / liquid oxygen and compressed hydrogen / compressed oxygen.

A polymer electrolyte membrane is an important component of a PEM fuel cell that is connected between the electrodes (anode and cathode). The cathode must be supplied by oxygen gas, whereas the anode must be supplied with hydrogen. The overall electrochemical dynamic can be represented by following equations [19]:

Cathode: $O_2 + 4H^+ + 4e^- \leftrightarrow 2H_2O$ (5)

Anode: $2H_2 \leftrightarrow 4H^+ + 4e^-$ (6)

Overall: $2H_2 + O_2 \leftrightarrow 2H_2O + electricity + heat$ (7)

For any fuel cell, both the anode and cathode can be represented by the mole conservation equations as follows [19]:

$$\frac{dP_{H2}}{dt} = \frac{RT}{V_a}[H_{2in} - H_{2used} - H_{2out}] \quad (8)$$

$$\frac{dP_{O2}}{dt} = \frac{RT}{V_c}[O_{2in} - O_{2used} - O_{2out}] \quad (9)$$

The fuel cell dynamic is built in a model in Simulink using a controlled voltage source in series with a constant resistance as illustrated in Fig. 10 [20].

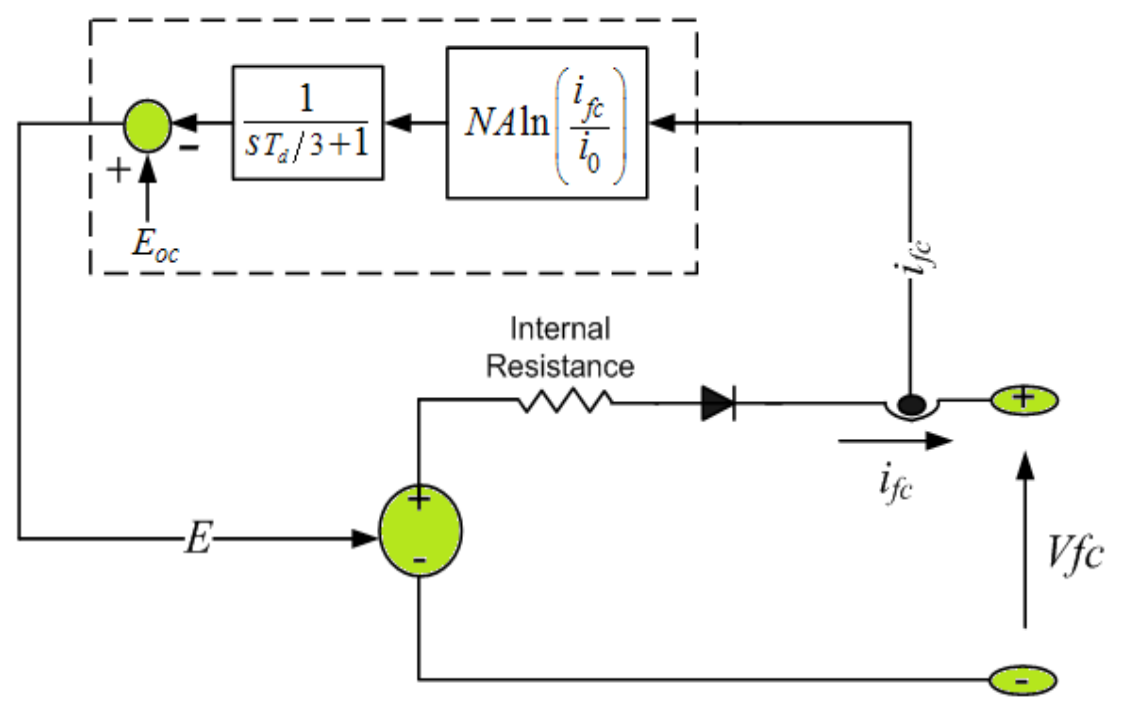

Fig. 10: Fuel cell stack model [19]

Equation (10) describes the controlled voltage source (E), so that

$$E = E_{oc} - NAln\left(\frac{i_{fc}}{i_0}\right) * \frac{1}{sT_d/3+1} \quad (10)$$

$$V_{fc} = E - R_{ohm} * i_{fc} \quad (11)$$

Equation (10) shows the fuel cell stack voltage as function of activation losses because of the slowness of chemical reactions at the electrode surfaces [20]. A parallel RC branch is used to model the losses electrically. Thus, for the rapid changes in the fuel cell current, the stack voltage will demonstrate a delay response that can be 3 times to the time constant $(\tau = RC)$ prior to equilibrium. Equation (10) also illustrates a phenomenon which delays the activation losses with a first order transfer function $\left(\frac{1}{sT_d/3+1}\right)$ where $T_d$ is the stack settling time. Equation (11) represents the total fuel cell voltage by taking the losses into account due to electrodes and electrolyte resistances (ohmic losses). This model is a simplified model that can simulate a fuel cell stack at a nominal condition of pressure and temperature operations. To eliminate the flow of negative current into the fuel cell, a diode is used [20]. Polarization curves (V-I and P-I) from the simulation and data sheet are presented in Fig. 11 and Fig. 12, respectively. The results from both MALAB / Simulink and the manufacturer's data sheet align well. The performance characteristics data of the stack are given for baseline operating conditions and defined at sea level and room ambient temperature. More information about the fuel cell is attached in the appendix.

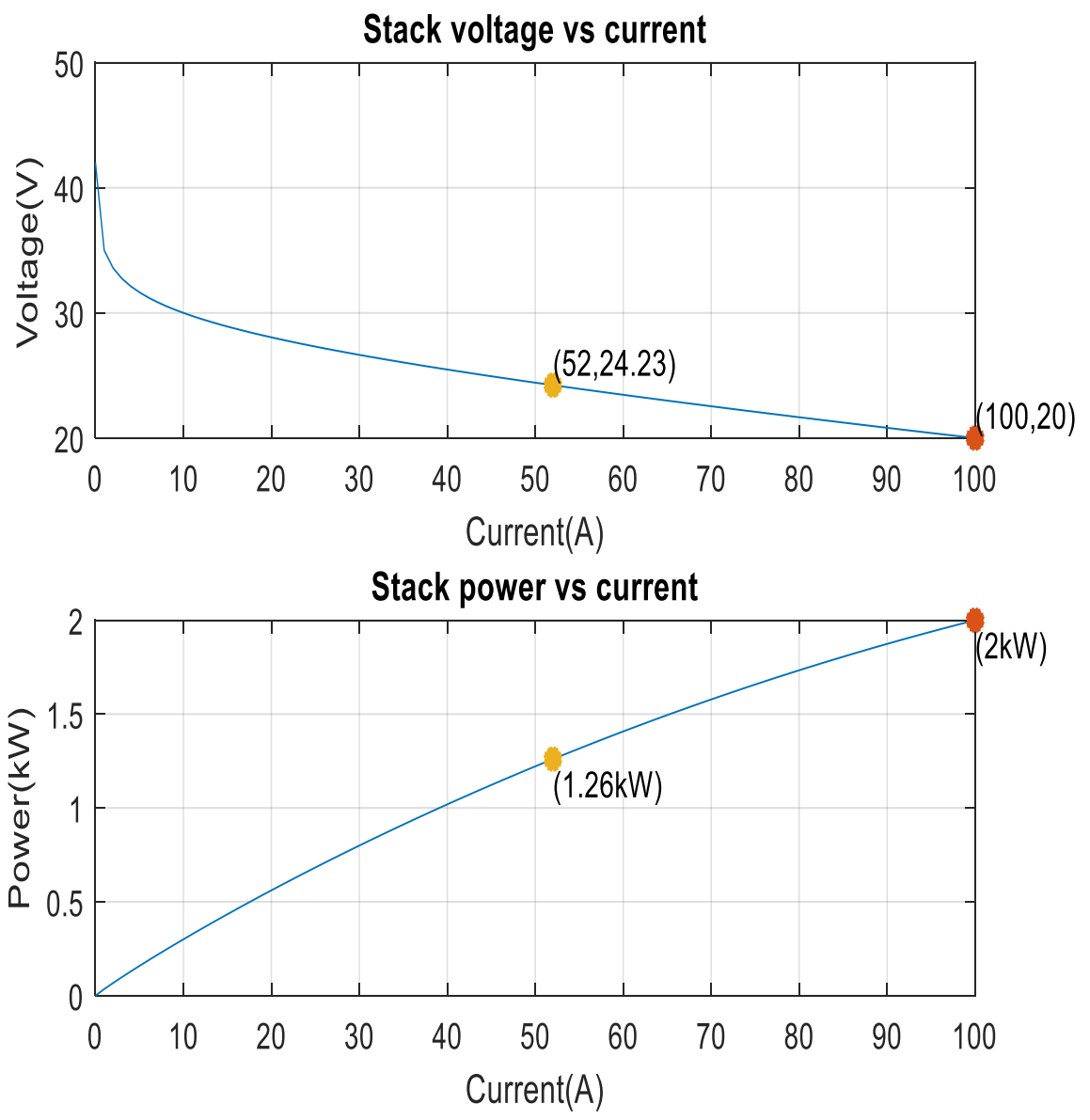

Fig.11: Polarization curves, voltage vs current and power vs current from simulation results

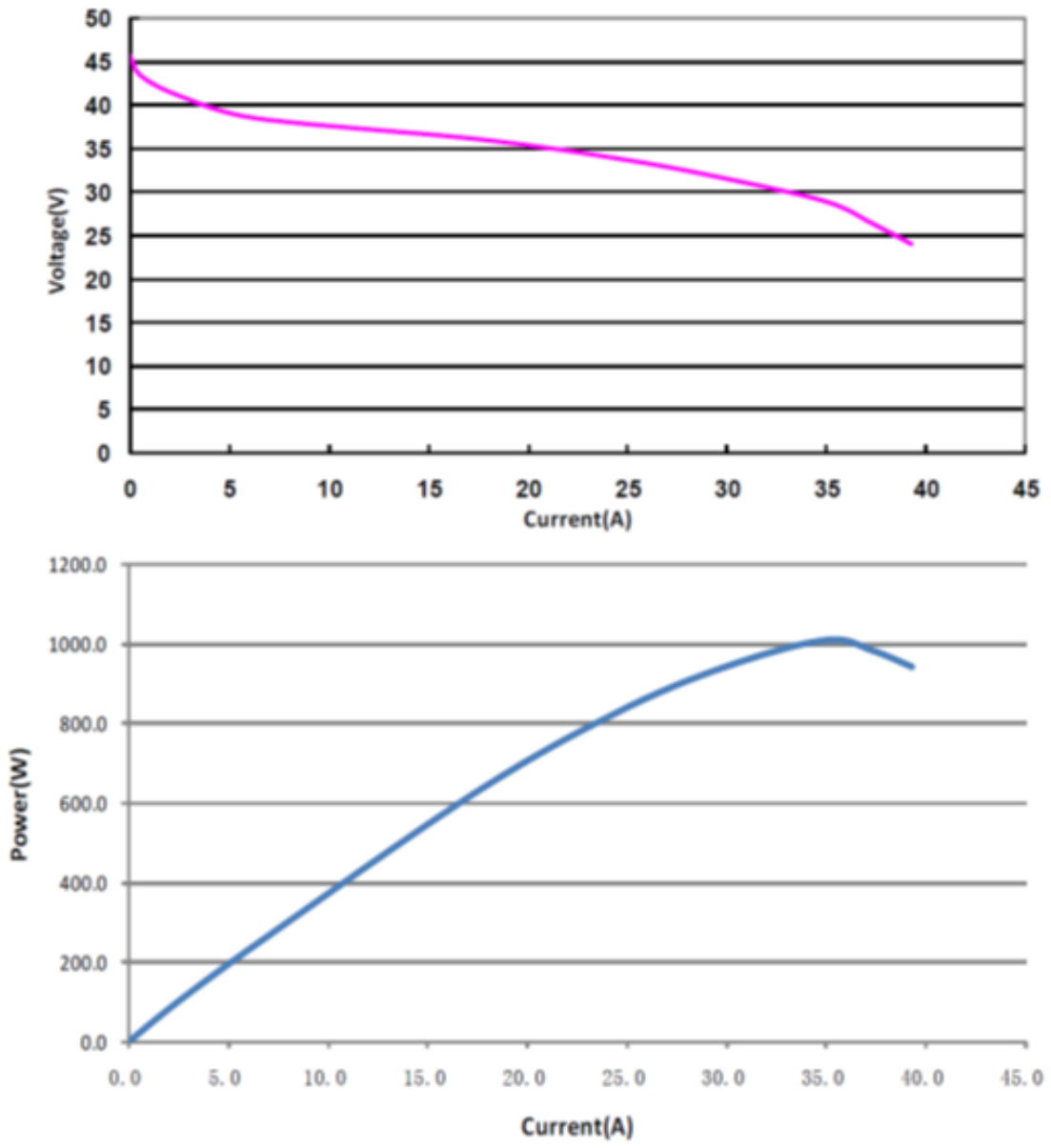

Fig.12: Polarization curves, voltage vs current and power vs current from data sheet results

### 2) Lithium-Ion Battery and Converter

MATLAB / Simulink already has a built-in dynamic model for a Li-ion battery that depends on a modified Shepherd curve-fitting model. The voltage polarization term was added to the battery discharge voltage expression to ensure the representation of the battery SOC effect on the battery performance. For the simulation stability, the filtered battery current is implemented instead of the actual battery current for the polarization resistance. The model uses two equations for discharging and charging as follows [21]:

*Discharge Model when $i^*$ is grater than Zero*

$$V_{batt} = E_0 - K\frac{Q}{Q-it} \cdot i^* - K \cdot \frac{Q}{Q-it} \cdot it + A \cdot exp(-B \cdot it) - R_b.I \quad (12)$$

*Charge Model when $i^*$ is less than Zero*

$$V_{batt} = E_0 - K.\frac{Q}{it+0.1Q} \cdot i^* - K \cdot \frac{Q}{Q-it} \cdot it + A \cdot exp(-B \cdot it) \quad (13)$$

Fig. 13 illustrates the dynamic model for a Li-ion battery in MATLAB / Simulink. Table 4 also shows the battery model input parameters.

Table 4: Battery model input parameters

| **Battery Model Input Parameters** | **Value** |
|---|---|
| **Nominal Voltage** | 48 (V) |
| **Rated capacity** | 34 (Ah) |
| **Maximum capacity** | 34 (Ah) |
| **Fully charged Voltage** | 55.87 (V) |
| **Nominal Discharge Current** | 14.78 (A) |
| **Internal Resistance** | 0.014(Ohm) |
| **Capacity at Nominal Voltage** | 30.74 (Ah) |

The simulation discharge curves for the Li-ion battery system (i.e. 48 V and 34 Ah) are shown in Fig. 14.

The average mode boost converter is used in this simulation, and its parameters are illustrated in Table 5. For the DC / DC converter parameters, some equations have been implemented to calculate the values for duty cycle (D), inductance (L), and capacitance (C) [22]:

$$D = 1 - \frac{(V_{in_min} * n)}{V_{out}} \quad (14)$$

$$L = \frac{(V_{in} * (V_{out} - V_{in}))}{(I_{in} * f_s * V_{out})}, and \quad (15)$$

$$C = \frac{I * D}{f_s * dv} \quad (16)$$

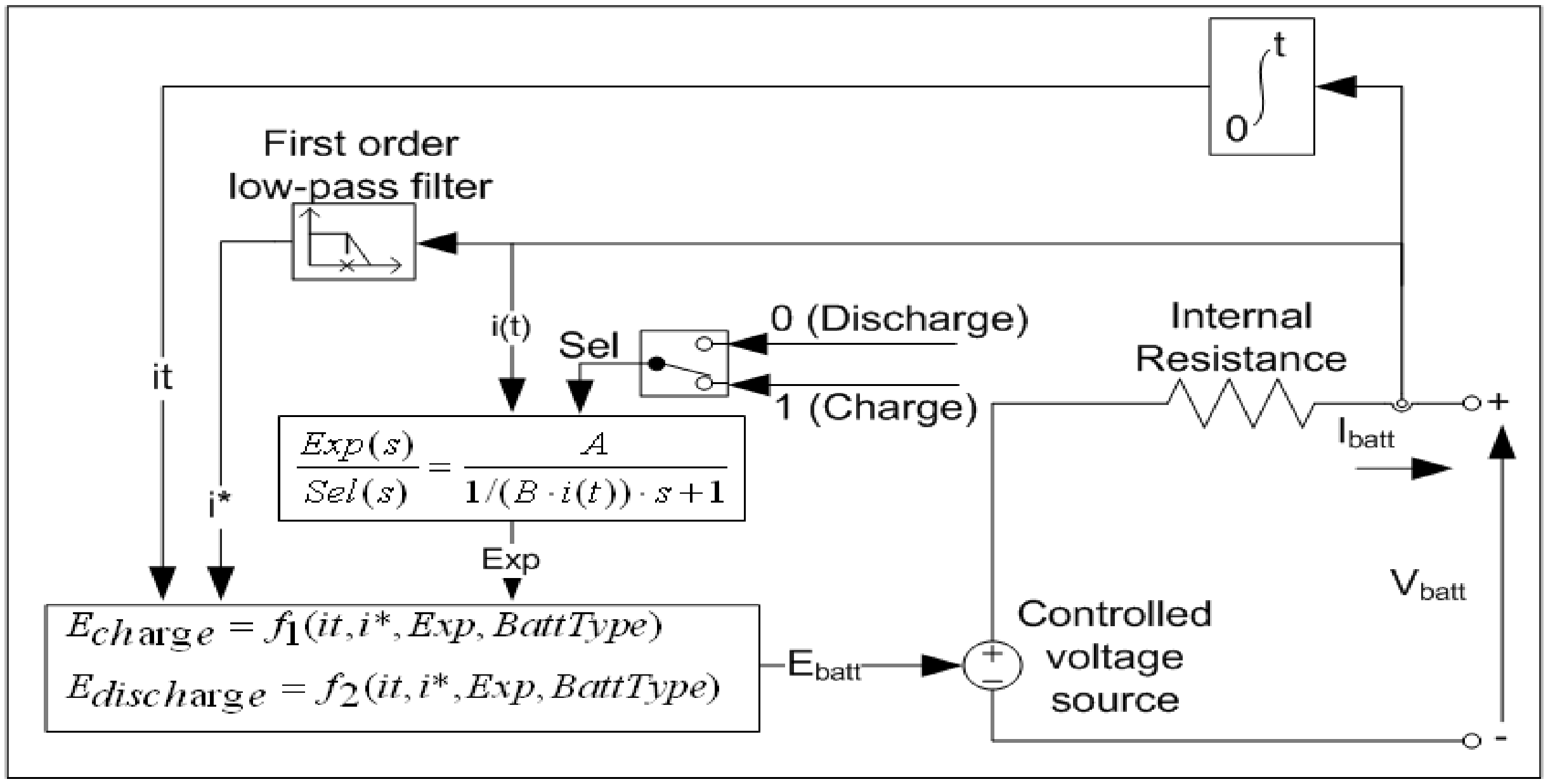

Fig. 13: Dynamic model for Li- ion battery [21]

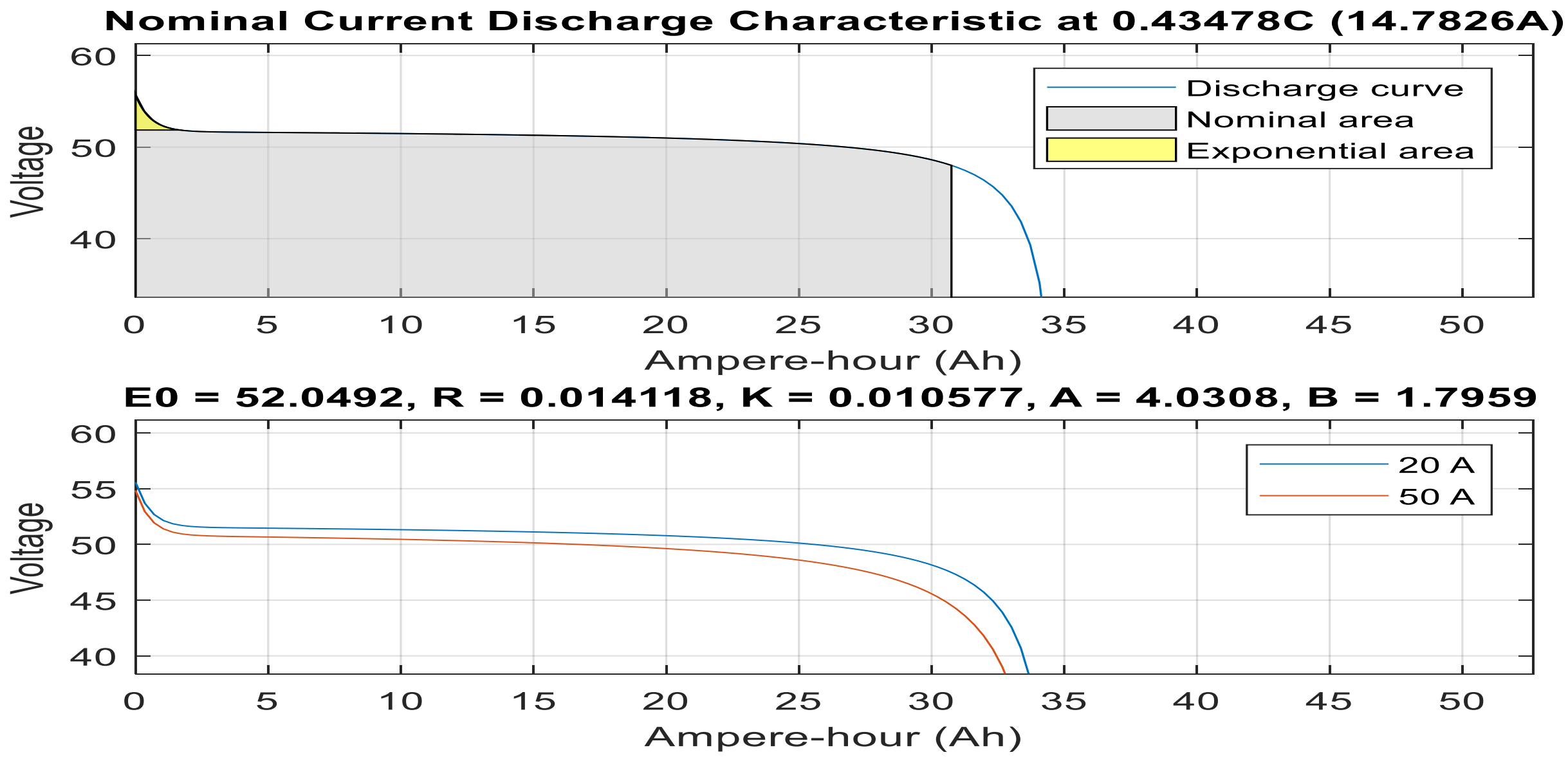

Fig. 14: Simulation discharge curves for the Li-ion battery

where D is the duty cycle, which equals to the fraction of time where the switch is connected in position 1, and hence $0 \leq D \leq 1$. $V_{in_min}$ is the minimum input voltage, n is the efficiency set to 90%. The variable $F_s$ is the switching frequency, $V_{out}$ is the output voltage, $I_{in}$ is the input current and dv is the output voltage ripple [23].

Table 5: Boost converter parameters

| Parameters | Value | Units |
|---|---|---|
| **Switching freq. F** | 20 | kHz |
| **Inductance L** | 500 | µH |
| **Capacitance C** | 7500 | µF |
| **Load Resistor R** | 0.2 | Ω |

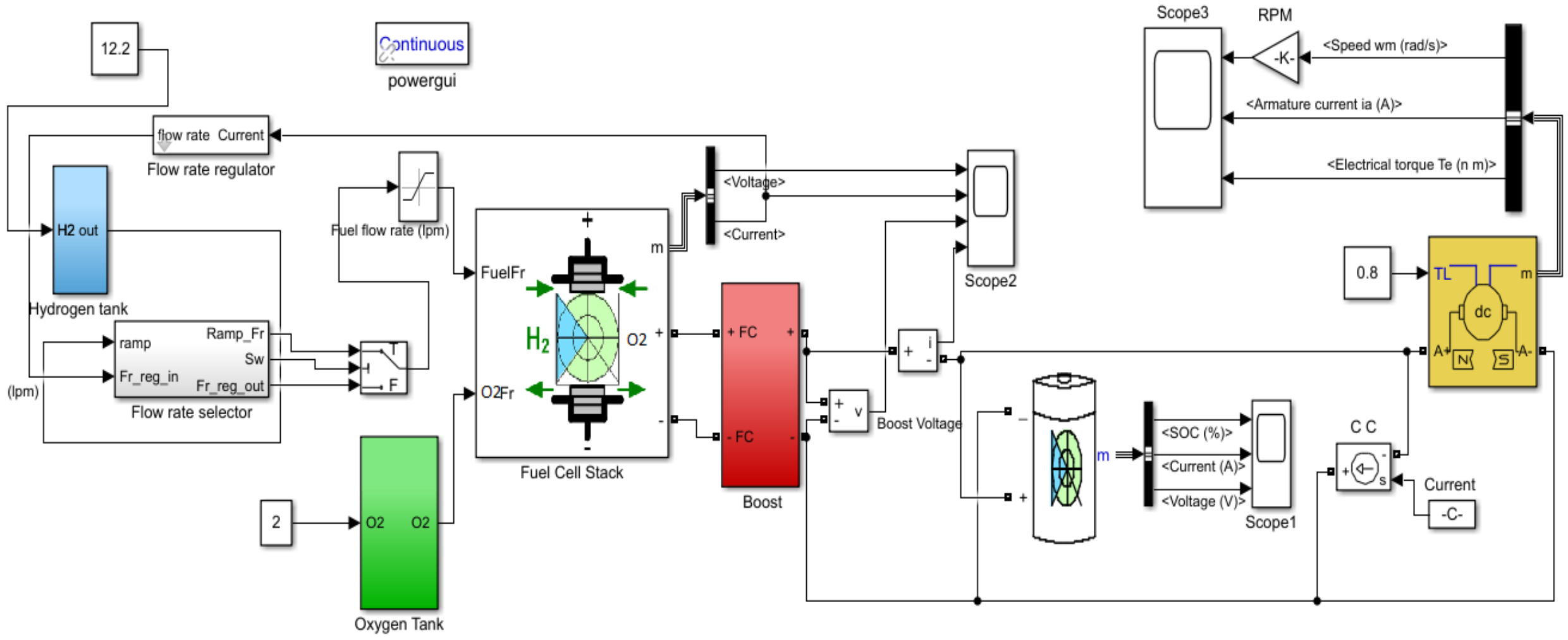

Fig. 15: Dynamic model in MATLAB/Simulink Software

### 3) Permanent Magnetic DC Motor (PMDC)

The dynamic model for any PMDC motor can be represented by the following equations [24]:

$$\frac{dI_a}{dt} = \frac{1}{L_{aa}} * (V_t - I_a * R_a - K_m * \omega_m) \qquad (17)$$

$$\frac{d\omega_m}{dt} = \frac{1}{J} * (T_e - T_L - B_m * \omega_m) \qquad (18)$$

Table 6 shows the parameters for the DC motor implemented in MATLAB / Simulink. Most of these values were collected from the DC motor datasheet. Fig. 15 shows the system dynamic flow rate regulators and flow rate selector. The blue and green blocks represent the hydrogen and oxygen tanks, respectively. They both enter the fuel cell stack in order to get power. The fuel cell is connected to the boost converter to increase the voltage from 24 V to 48 V, which is required by the battery and the load (i.e. DC motor). The yellow block illustrates the MUN Explorer’s motor.

Table 6: PMDC motor parameters

| **Parameters** | **Value** | **Unit** |
|---|---|---|
| **Armature V** | 48 | V |
| **Armature Ra** | 0.3 | Ohms |
| **Armature La** | 0.00208 | H |
| **Torque constant** | 0.099 | N.m/A |
| **Total Inertia J** | 15e-5 | Kg.m^2 |

## IV. Results and Discussion

The simulation in HOMER software was done to get the sizing results for the integrated power system. The system component inputs were specified based on the cost and sizes to consider for each block. From Fig. 2, the components of the PV wind turbine and electrolyzer can not be applied to the MUN Explorer due to the lack of space available, and they will be used to generate the required oxygen and hydrogen gases to run the fuel cell. From Fig. 4, the assumptions of the wind and PV energy are determined based on the wind speed directions and solar radiations of St. John’s, Newfoundland, which is where this technology will be integrated.  The oxygen /

hydrogen tanks and fuel cell along with the batteries are planned to be installed in the MUN Explorer. The gray line also shows the most optimal results. From my point of view, the results have shown the lowest operating cost and reduce the number of batteries from 11 to 8. The advantage of minimizing the number of batteries is that it leaves more space for installing the fuel cell and the tanks. The hydrogen stored in the cylinders can be generated from renewable energy sources as a step prior to running the MUN Explorer. Figures 16 and 17 show the pressure inside the oxygen and hydrogen tanks, respectively,

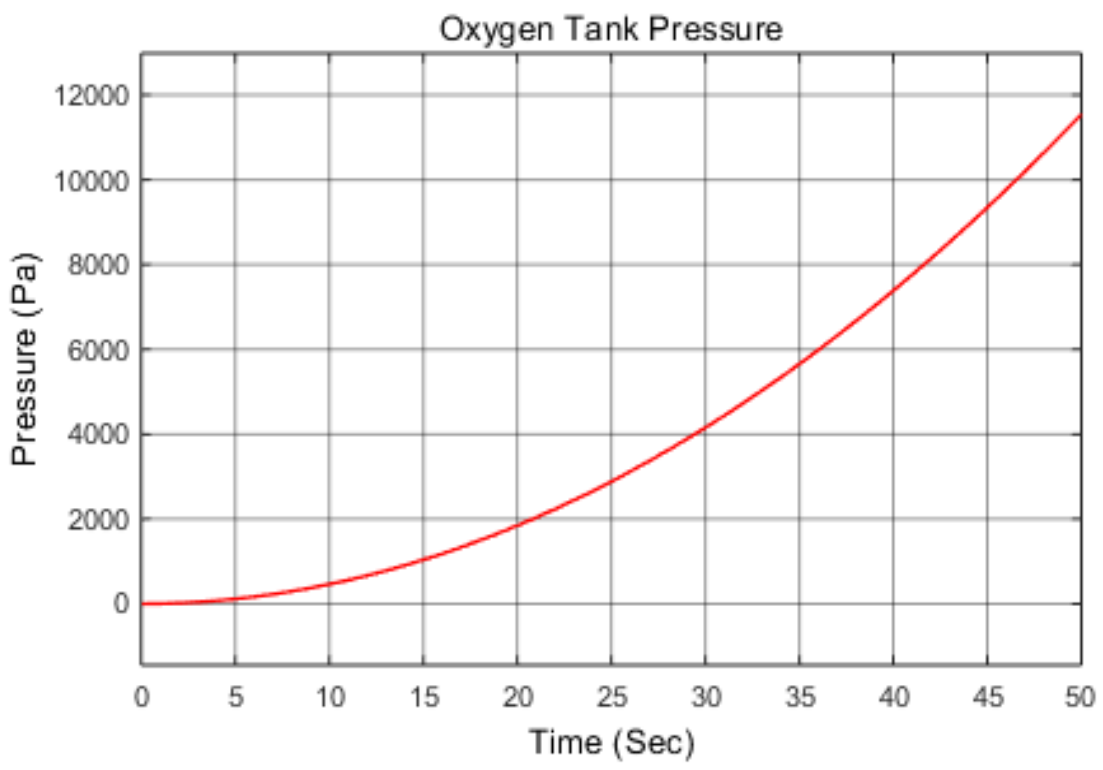

Fig. 16: The pressure of compressed oxygen tank

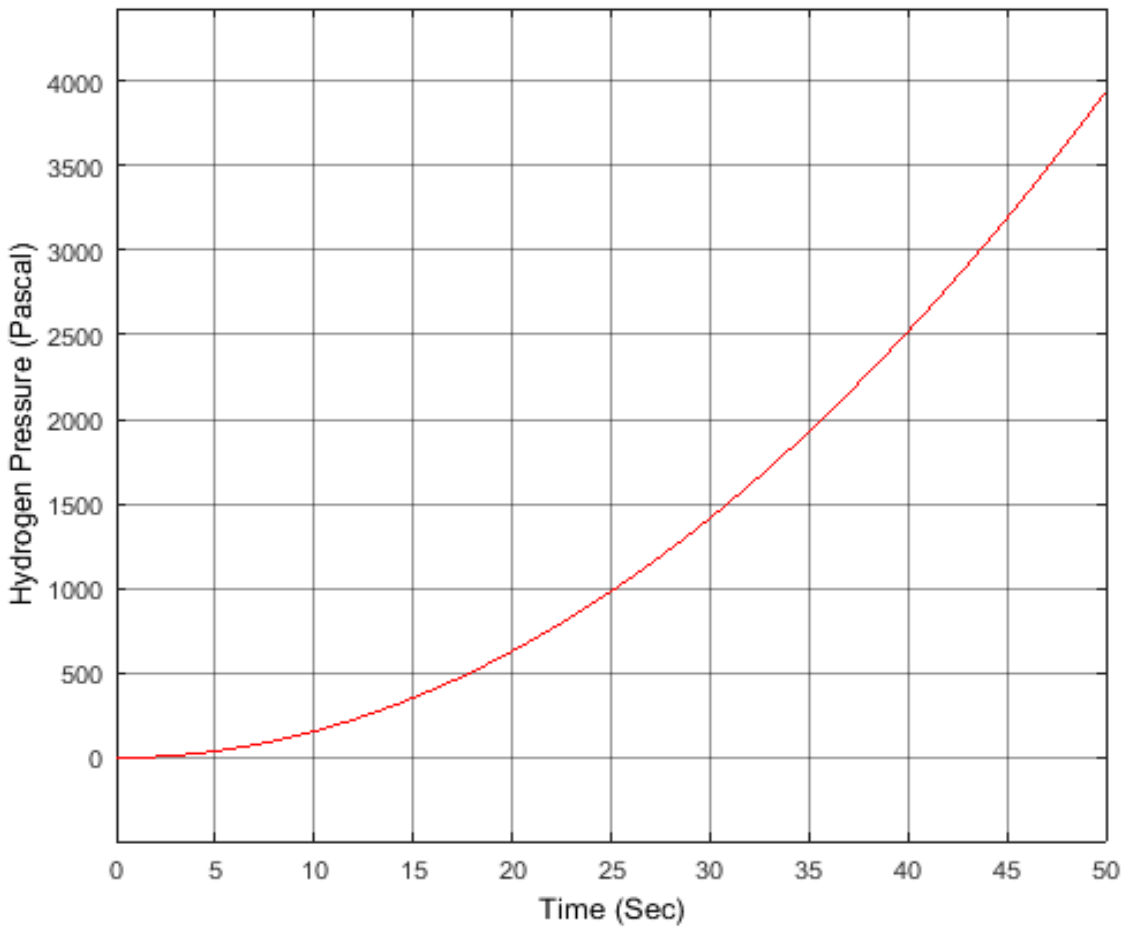

Fig. 17: The pressure of compressed hydrogen tank

A 6-kg hydrogen tank needs to be installed inside the explorer to run the fuel cell. In terms of the oxygen tank, it can be installed as the hydrogen tank specifications which was suggested in the literature. The oxygen tank size is considered the same as the one in [16]. The fuel cell has 1 kW of power to feed the DC motor and to charge the battery once it gets enough power. Firstly, the fuel cell is connected to the boost converter, which takes 24 V to 48 V, which is required by the battery and DC motor, as shown in Fig. 18. The assumptions of the fuel cell model are: all gases are ideal, pressure drops across flow channels are negligible and cell voltage drops are due to reaction kinetics and charge transport [19]. A PI controller is used to control the output voltage from the boost converter to maintain the 48 V for the battery and DC motor. The PI coefficients are shown in Table 7.

Table 7: PI coefficients for boost converter

| Parameter | Value |
|---|---|
| **$K_P$** | 0.0005 |
| **$K_i$** | 0.15 |

From Fig. 18, we can clearly see that after 20 seconds, the fuel cell started to run to power the DC motor. This starting time is recommended by the fuel cell manufacturer and controlled by the fuel cell regulator. The battery is set to 50% state of charge (SOC) to prevent any damage to the battery and does not allow it to charge to 100%. The nominal discharge current is 14.78 A. Fig. 19 shows the fuel cell power profile in HOMER through the year.

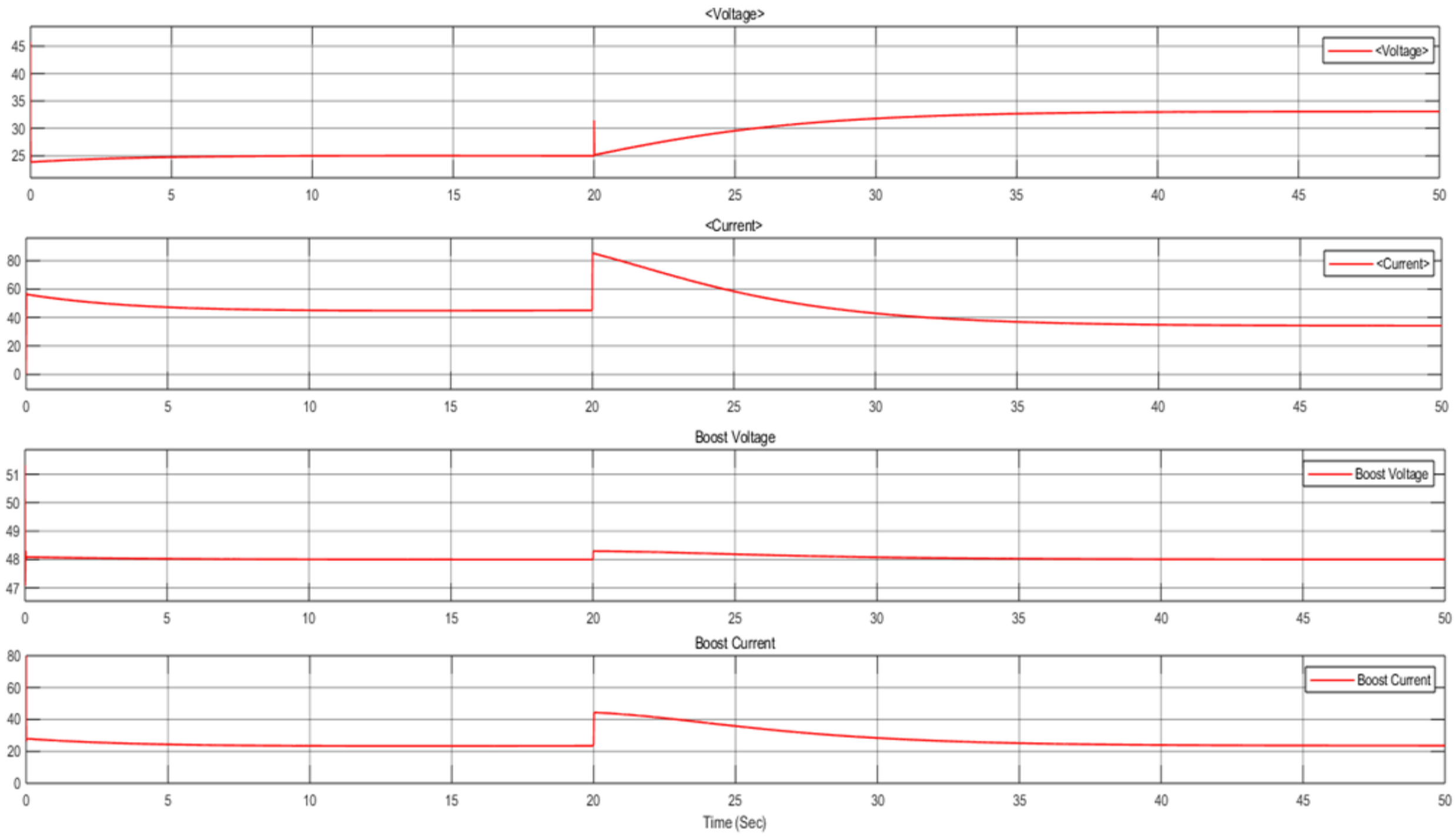

Fig. 18: The voltage and current of fuel cell and Boost converter

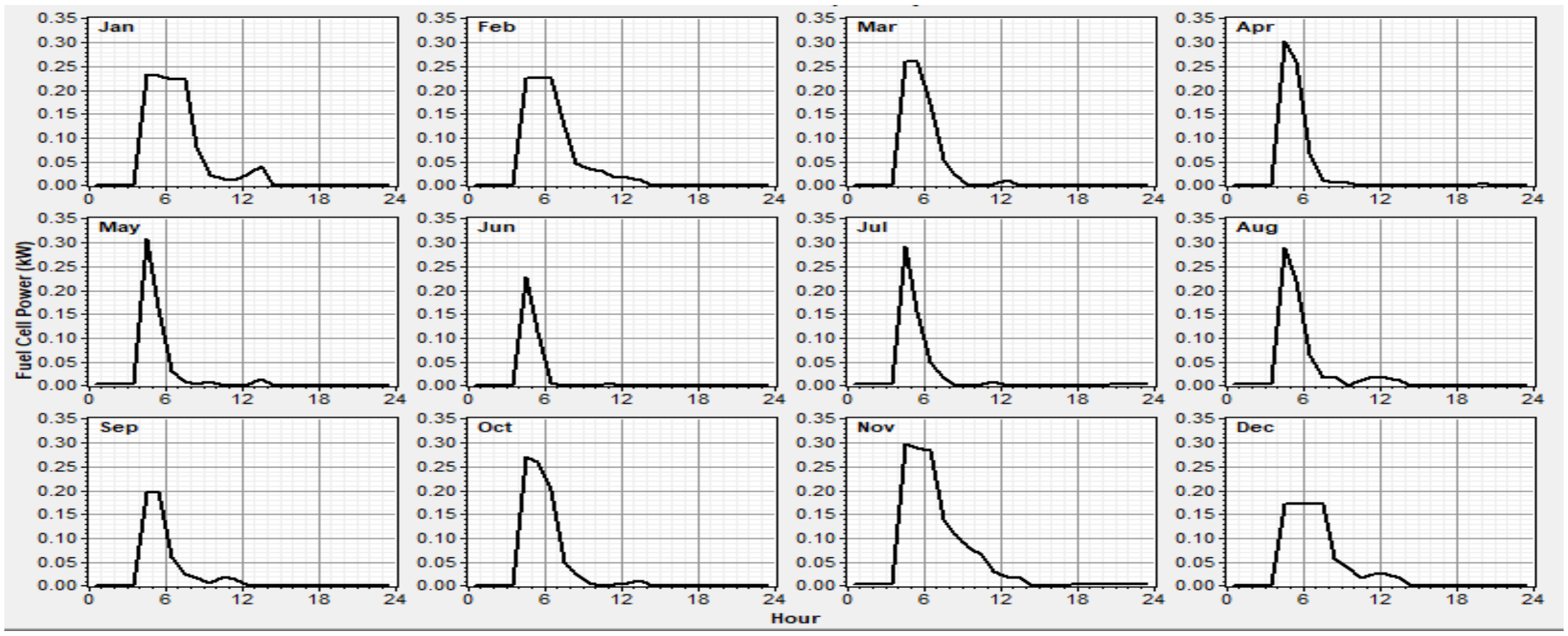

Fig. 19: The fuel cell power profile

There are 11 batteries connected in series to power the AUV for 10 hours. Fig. 20 illustrates the battery behavior in terms of SOC, current and voltage from the Simulink model, whereas Fig. 21 shows the state of charge profile from HOMER sizing during the year. Fig. 22 illustrates the PMDC motor, which runs at a constant speed. This constant speed is maintained by the boost converter to run the AUV at a constant speed. After that, the DC motor runs at its highest efficiency. The armature current is 16, which is very close to the manufacturer data sheet value. Fig. 23 demonstrates the DC motor power profile in HOMER software.

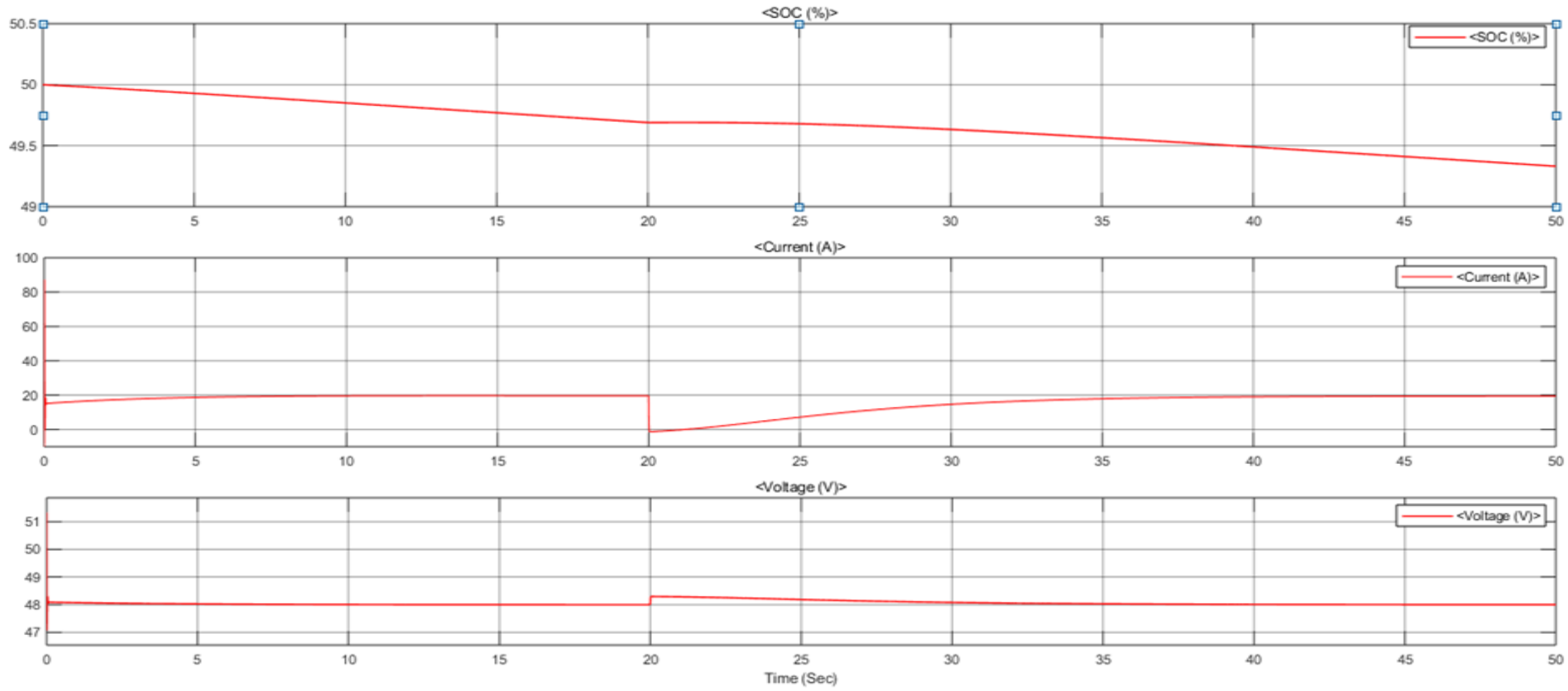

Fig. 20: The battery characteristic SOC, current and voltage

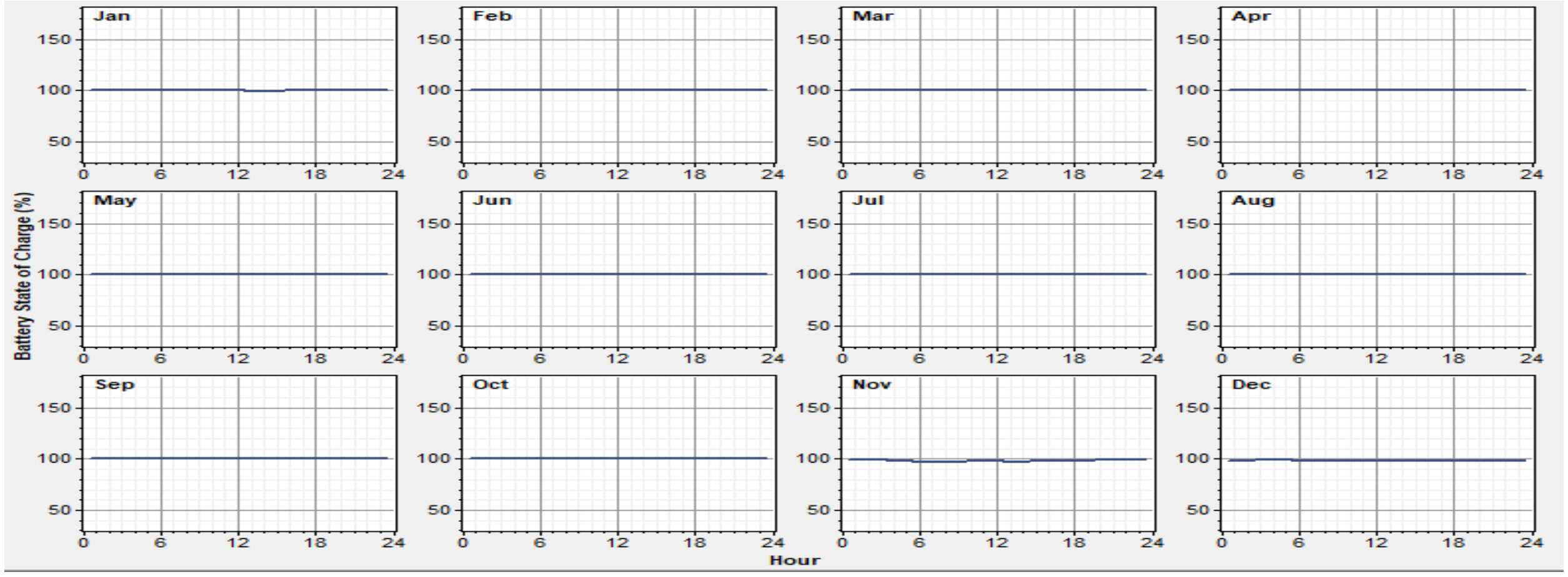

Fig. 21: The battery power profile

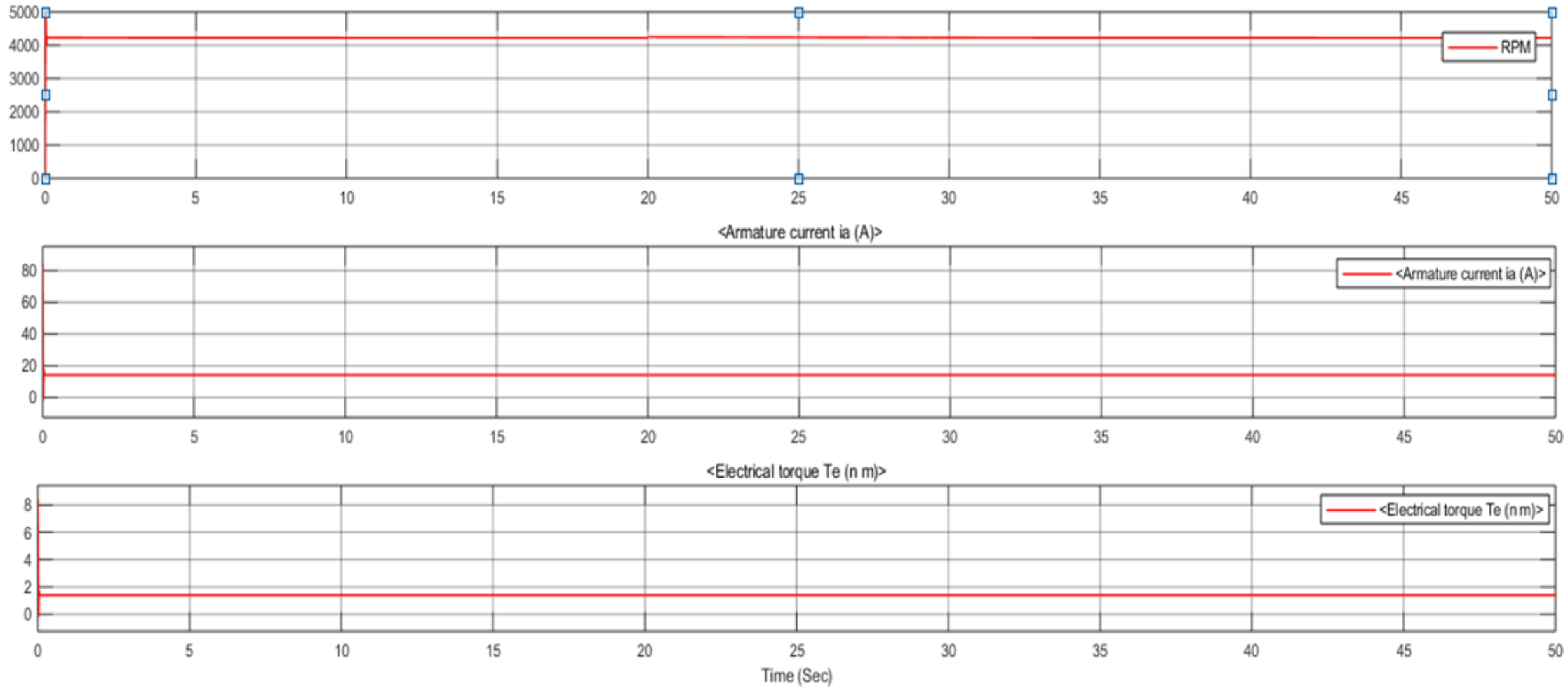

Fig. 22: PMDC motor characteristics

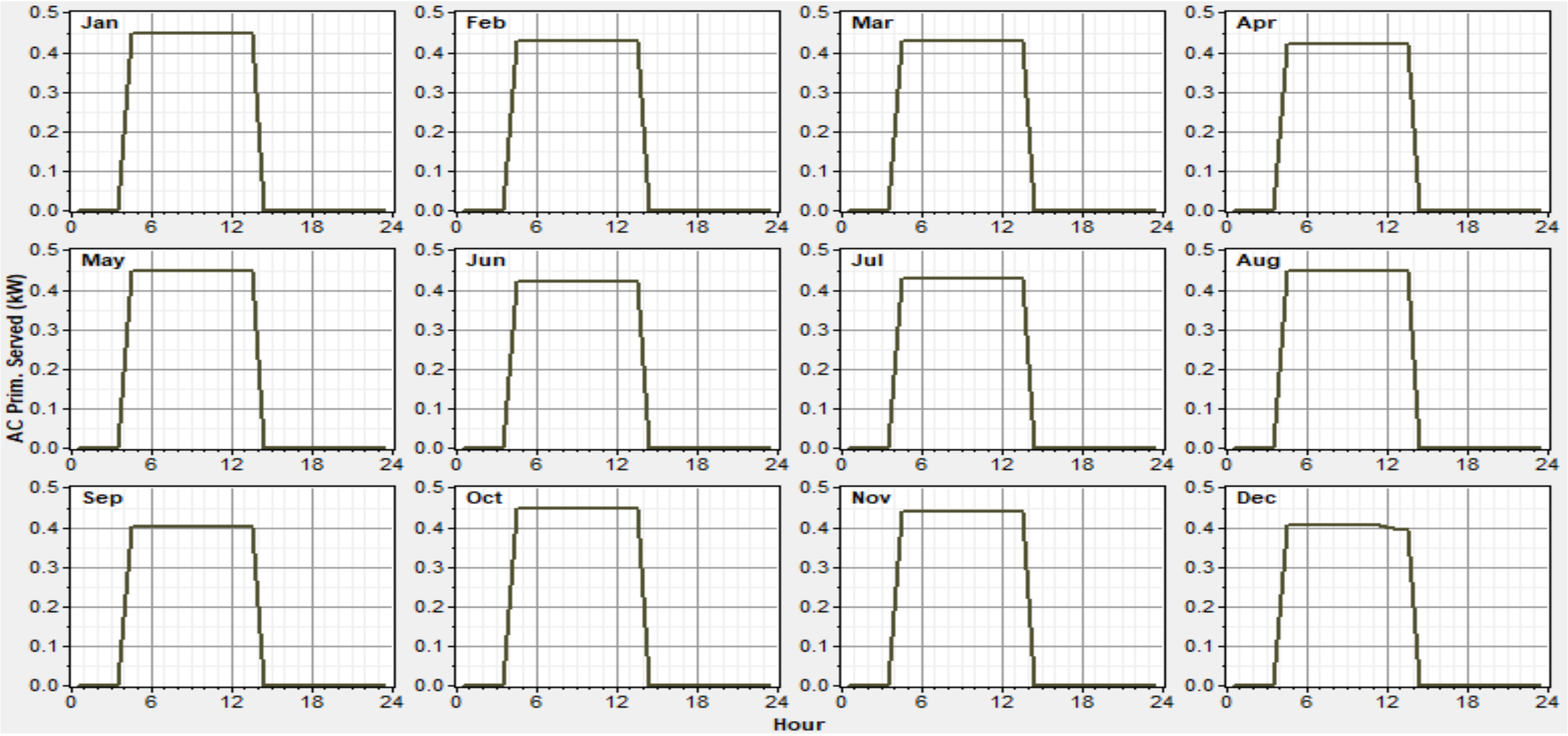

Fig. 23: The DC motor power profile

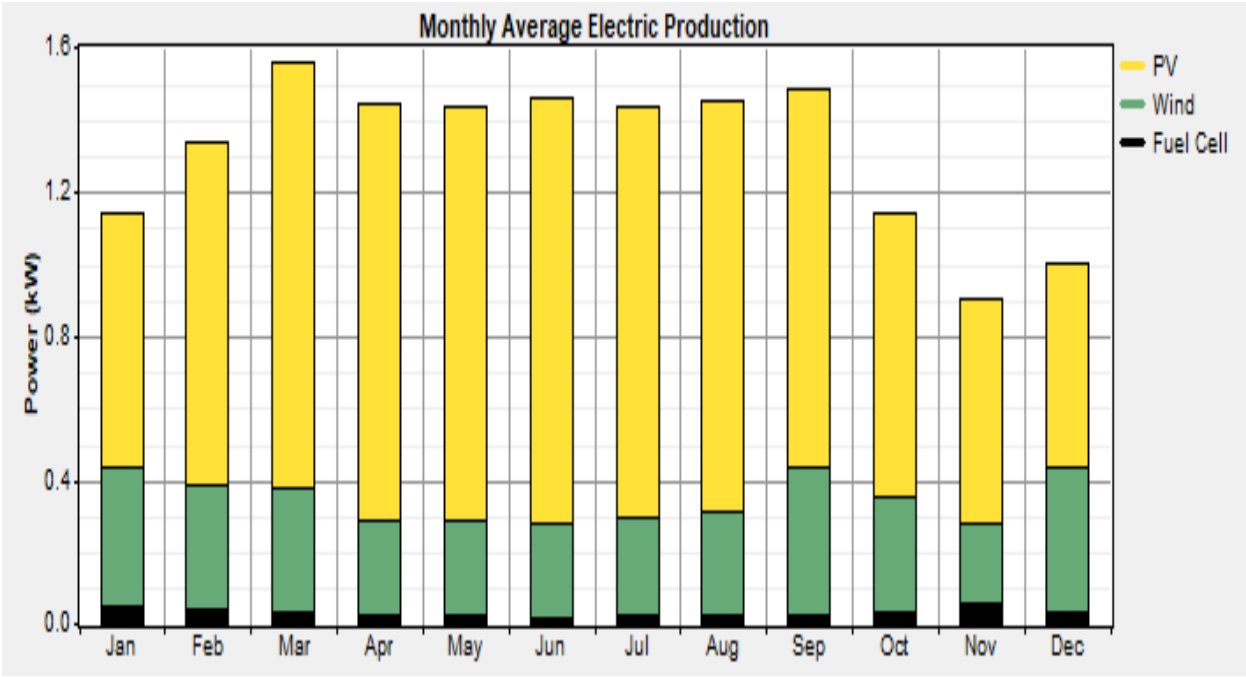

Fig. 24: Monthly average electric production

The monthly average electric power production from the system (i.e. PV, wind and fuel cell) is shown in Fig. 24. The solar and wind energy are used to produce oxygen and hydrogen gas by powering an electrolyzer as well as recharging the batteries. The fuel cell power production is low compared to the PV and wind power due to the integration of the fuel cell into the battery system, which has a large amount of energy to power the DC motor. Table 8 shows the results for energy density and specific energy for the storage and total system with the fuel cell. The calculations are based on the total system's mass and weight, which can be illustrated as summations of the storage and fuel cell systems. The estimated results are shown due to the main balance of plant components that were integrated into commercial fuel cells.

Table 8: SE and ED for storage and total fuel cell

| | **Storage System** | | **Total system** | |
|---|---|---|---|---|
| | SE (kWh/kg) | ED (kWh/L) | SE (kWh/kg) | ED (kWh/L) |
| **$L_{O2}/L_{H2}$** | 1.233 | 1.130 | 1.024 | 0.846 |
| **$C_{O2}/C_{H2}$** | 0.792 | 0.379 | 0.701 | 0.339 |

Table 8 shows a significant improvement in terms of specific energy and energy density, especially for liquid oxygen and hydrogen storage options. In [12], Li-ion batteries have specific energies from 0.165 kWh/kg to 0.207 kWh/kg and energy densities from 0.329 kWh/L to 0.490 kWh/L [12]. The largest improvements are in the specific energy of the fuel cell total systems when compared with the lithium-ion batteries. To show the buoyancy effect on the system, the density can be defined as energy density divided by specific energy [14]:

$$D = \frac{m}{V} = \frac{ED}{SE} \quad (19)$$

Fig. 25 shows the relationship between ED and SE, the plotting of ED as functions of SE on the X and Y axes is the slope (i.e. X and Y intercept at any point) which is equivalent density at any point. The seawater density (1.03 kg/L) is shown by the dotted line. However, if there is any point above the line, it indicates negative buoyancy or denser than seawater. Any point below the line indicates positive buoyancy or less dense than seawater. For any given fuel cell power system design that does not require buoyancy as shown in Fig. 25, some ballast or float material must be added to meet the buoyancy requirement [14].

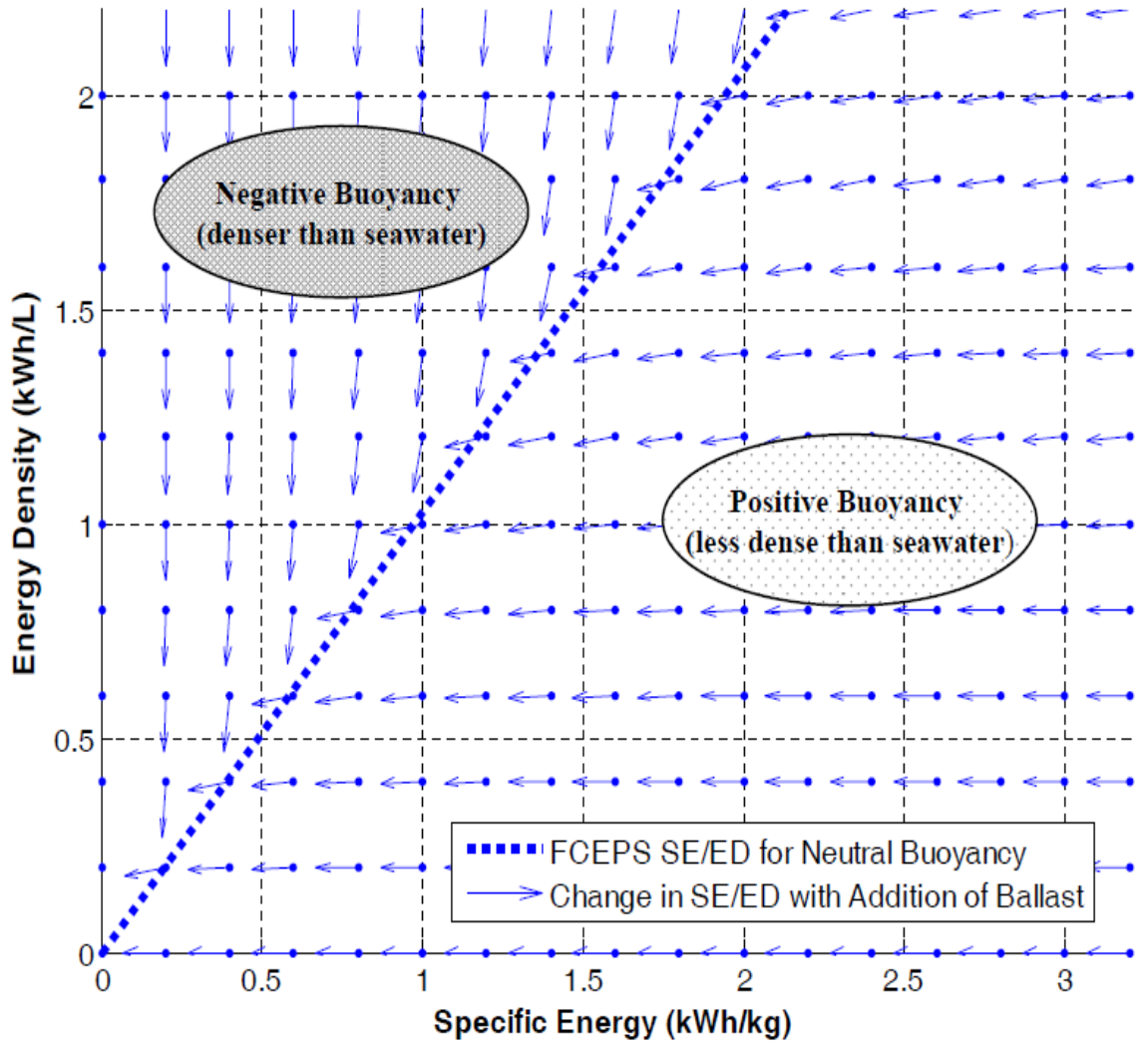

Fig. 25: Buoyancy in terms of SE and ED [14]

The power system's capacity is increased by integrating the fuel cell power system into the MUN Explorer according to the following calculations.

Available Energy=Power*time (20)

Watt-hour=Battery Volt*Ah (21)

The energy capacity is increased by integrating the fuel cell into the system and the number of batteries is reduced by applying equations 20 and 21. More details are attached in the appendix.

## V. Conclusions

The sizing and modeling of the MUN Explorer's power system were studied and simulated in this paper. The oxygen and hydrogen tanks were successfully studied in terms of specific energy and energy density. They were also implemented in MATLAB / Simulink as compressed gas storage. The results showed that a fuel cell with hydrogen and oxygen storage options has a higher energy density than batteries alone. The system sizing by HOMER was studied and implemented. The power profiles from HOMER software were illustrated for the fuel cell and DC motor. A 1-kW fuel cell and 8 Li-ion batteries can increase the power system capacity to 68 kWh. Installing these options will greatly increase the hours of operation and will help the buoyancy force.

The system components are simulated in MATLAB / Simulink.

Future work that builds on this paper should improve the dynamic model in MATLAB / Simulink by including some controllers in the system. The existing power system for the MUN Explorer should be built and compared with this system (i.e. fuel cell with batteries).

## Nomenclature

AUV: Autonomous underwater vehicle
DC: Direct current.
Pb: Pressure of the tank (Pa)
$P_{bi}$: Initial pressure of the tank (Pa)
$T_b$: Operating temperature (K)
$N_{H2}$: Normal hydrogen flow rate (Liter/min)
$V_b$: Volume of the tank ($m^3$)
T: Temperature (K)
Z: Compressibility factor
$V_m$: Molar volume ($m^3$)
P: Pressure (pa)
ED: Energy density (kWh/L)
SE: Specific energy (kWh/kg)
SS: Storage system
$H_2O$: Water
$H_2$: Hydrogen gas
$O_2$: Oxygen gas
$P_{H2}$: Hydrogen pressure Anode side (Pa)
R: Universal gas constant (J/ (mol.K)
$V_a$: Anode's volume (m3)
$H_{2in}$: Hydrogen input flow rate (kg/ sec)
$H_{2out}$: Hydrogen output flow rate (kg/ sec)
$P_{o2}$: Oxygen pressure cathode side (Pa)
$V_c$: Cathode's volume ($m^3$)
$O_{2in}$: Oxygen input flow rate (kg/ sec)
$O_{2out}$: Oxygen output flow rate (kg/ sec)
E: Controlled voltage source (V)
$E_{OC}$: Open circuit voltage (V)
N: Number of cells
A: Tafel slope (V)
$i_0$: Exchange current (A)
$T_{d:}$ Response time (sec)
$R_{ohm}$: Internal resistance (ohm)
$i_{FC}$: Fuel cell current (A)
$V_{FC}$: Fuel cell voltage (V)
τ: Constant time
$E_{Batt}$: Nonlinear voltage (V)
$E_0$: Constant voltage(V)
Exp(s): Exponential zone dynamics (V)
K: Polarization constant ($Ah^{-1}$)
i*: Low frequency current dynamics(A)
i: Battery current(A)
it: Extracted capacity (Ah)
Q: Maximum battery capacity (Ah)
A: Exponential voltage(V)
B: Exponential capacity ($Ah^{-1}$)
Ah: Ampere hour
COE: Levelized cost of energy ($/kWh)
NPC: Total net present cost of a system
D: Duty cycle
$V_{in_min}$: Minimum input voltage(V)
n: Efficiency
$F_s$: Switching frequency(V)
$V_{out}$: Output voltage(V)
$I_{in}$: Input current(A)
dv: Output voltage ripple(V)
L: Inductance (H)
C: Capacitance (F)
$V_t$: DC source voltage (V)
$I_a$: Armature current (A)
$R_a$: Armature resistance (Ω)
$L_{aa}$: Armature inductance (H)
J: Inertia constant (kg*m2)
$B_m$: Constant (N ∗m∗s)
$K_m$: Torque constant (V·s/rad)
$\omega_m$: Motor speed (rpm)
D: Density (kg/L)
V: Volume ($m^3$)
M: Mass (kg)

## Conflict of Interest

The authors declare that they do not have any conflict of interest in regard to this publication.

## Acknowledgments

The authors would like to thank Libyan Government for financial support for this work.

## Data Availability Statement

The data used to support the findings of this study are available from the corresponding author upon request.

# Appendix

| **Components** | **Cost** |
|---|---|
| Fuel cell (500/1k) W | $3084/ $4284 |
| Electrolyzer | $1509 |
| Hydrogen Tank | $915 |
| Battery | $840 |
| Wind turbine | $800 |
| PV panel | $3600 |
| DC Motor | $60 |

## Specifications

| Characteristics | Specifications |
|---|---|
| Length | 5.3 m |
| Diameter | 0.69 m |
| Dry Weight | 830 kg |
| Energy | 17.6 kWh |
| Maximum Depth | 3000 m with 10% safety factor |
| Typical Crusing Speed | 1.5 m/s |
| Speed Range | 0.5 m/s to 2.5 m/s |
| Power Source and Capacity | 11 x 1.6 kWh E-One Moli Energy Li-Ion Cobalt rechargeable battery modules |
| Computer | Rack mount cPCI system for vehicle control and payload control computer |
| Hydroplanes | 4 NACA 0026 stern planes<br>2 NACA 0026 fore planes |
| Navigation INU Type | iXsea PHINS III |

https://www.mun.ca/engineering/research/facilities/centres/oerc/facilities/merlin/explorerauv.php

## Technical data

**E206:**

Item name: ........................ Electrolyser H2/O2 65
Item no: ........................................................ E206
H x W x D: ........................... 250 x 250 x 120 mm
Weight: ........................................................ 950 g
Number of cells: ................................................ 2
Electrode dimensions: ........................ 40 x 40 mm
Operating medium: .... distilled water, $\sigma < 2$ µS/cm
Fill volume H2O, H2-side: ............... approx. 90 ml
Fill volume H2O, O2-side: ............. approx. 130 ml
Permissible operating voltage: ........... 0 - 4.0 VDC
Permissible operating current: ............... 0 - 4.4 A
Rated power consumption: ............. approx. 16 W
Gas production H2 at rated power output: ... approx. 65 $cm^3$/min
Gas production O2 at rated power output: ... approx. 32.5 $cm^3$/min

**E207:**

Item name: ...................... Electrolyser H2/O2 230
Item no: ........................................................ E207
H x W x D: ........................... 250 x 330 x 200 mm
Weight: ...................................................... 1850 g
Number of cells: ................................................ 7
Electrode dimensions: ........................ 40 x 40 mm
Operating medium: .... distilled water, $\sigma < 2$ µS/cm
Fill volume H2O, H2-side: ............... approx. 90 ml
Fill volume H2O, O2-side: ............. approx. 130 ml
Permissible operating voltage: ......... 0 - 14.0 VDC
Permissible operating current: ............... 0 - 4.4 A
Rated power consumption: ............. approx. 56 W
Gas production H2 at rated power output: ... approx. 230 $cm^3$/min
Gas production O2 at rated power output: ... approx. 115 $cm^3$/min

| Fuel Cell Properties | |
|---|---|
| Number of Cells | 48 |
| Rated Power | 1000W (1kW) |
| Rated Performance | 28.8V @ 35A |
| Hydrogen Supply Valve Voltage | 12V |
| Purging Valve Voltage | 12V |
| Blower Voltage | 12V |
| Reactants | Hydrogen and Air |
| Ambient Temperature | 5 - 30C (41 - 86F) |
| Max Stack Temperature | 65 C (149 F) |
| Hydrogen Pressure | 0.45 - 0.55 Bar |
| Humidification | Self-humidified |
| Cooling | Air (integrated cooling fan) |
| Controller Weight | 400g (± 30g) |
| Stack Weight (with Fan & Casing) | 4kg ± 100g |
| Hydrogen Flow Rate at Max Output | 13 L/min |
| Stack Size | 268 x 219 x 122.5mm (10.5" x 8.6" x 4.8") |
| Hydrogen Purity Requirement | ≥ 99.995% (dry H2) |

https://www.fuelcellstore.com/fuel-cell-stacks/high-power-fuel-cell-stacks/horizon-1000watt-fuel-cell-h-1000

EXIDE TECHNOLOGIES

**ONYX +48V Battery Module User Manual**

170-400 R01.06

### 2.3 Module Electrical Specification

| **Capacity** | |
|---|---|
| ONYX +48V (per module nominal) | |
| M70X48V034P | 34Ah |
| **Module Charge** | |
| $V_{min}$ | +42VDC |
| $V_{max}$ = Charge Voltage | +54.0VDC |
| **Current** | |
| $I_{discharge(peak: less than 30 seconds)}$ | -20A |
| $I_{discharge(continuous)}$ | -15A |
| $I_{charge(peak: less than 30 seconds)}$ | +20A |
| $I_{charge(continuous)}$ | +15A |
| **Discharge Temperature Range** | |
| $T_{min(Discharge)}$ | -20℃ (-4℉) |
| $T_{max(Discharge)}$ | 60℃ (140℉) |
| **Charge Temperature Range** | |
| $T_{min(Charge)}$ | 0℃ (32℉) |

https://www.fuelcellstore.com/hydrogen-equipment/electrolyzer-230-e107

https://www.alibaba.com/product-detail/cheapest-1000W-wind-alternator-48v-for_60158460072.html

https://www.alibaba.com/product-detail/48v-brushless-dc-motor-nema34220w_60500232517.html?spm=a2700.7724838.2017115.96.57493907yRTCES

| Metal Hydride Properties | |
|---|---|
| Hydrogen Capacity - Alloy A | 34 standard liters (1.3 scf)* |
| Hydrogen Capacity - Alloys L, M, or H | 30 standard liters (1.14 scf)* |
| Hydrogen Pressure when Charging or Discharging | The hydrogen pressure when charging or discharging a SOLID-H™ container is something you select when you order. Four standard pressure ranges are offered; Alloy A (1-10 bar at room temperature), Alloy L (2-3 bar at room temperature), Alloy M (4-5 bar at room temperature) and Alloy H (8-12 bar at room temperature). |
| Discharge Rate | The discharge rate depends on many variables. We can help you select a SOLID-H™ metal hydride alloy and hydrogen container(s) that will meet your hydrogen flow requirements. In general, you should not expect to empty the entire hydrogen capacity in a matter of minutes. Hours are required to withdraw 90% or more of the hydrogen capacity from a standard** metal hydride container. The largest SOLID-H™ containers require days to discharge completely. **It is possible to discharge a metal hydride in a matter of seconds. This requires extraordinary heat transfer enhancement inside and outside of the container. We can provide heat transfer enhancement to improve the charging and discharging rates of our SOLID-H™ containers. |
| Recharge Time | About 4 hours - The specified recharge time is for cooling by still air at 20°C and the charging pressure specified in the SOLID-H™ manual for Alloys A, L, M or H. A fan will shorten charging time. |
| Cylinder Diameter | 1.125 inch (28.6 mm) |
| Overall Length | 7.8 inch (198 mm) |
| Mass | 0.96 lb (438 grams) |
| Destructive Proof Test | >5000 psig (350 bar) |
| Pressure Relief Valve Set | <550 psig (37 bar) |
| Materials Included | Stainless Steel Cylinder with Brass Fittings |

https://www.fuelcellstore.com/bl-30-metal-hydride

Fuel cell nominal parameters:
Stack Power:
-Nominal = 1259.96 W
-Maximal = 2000 W
Fuel Cell Resistance = 0.061871 ohms
Nerst voltage of one cell [En] = 1.115 V
Nominal Utilization:
-Hydrogen (H2)= 99.92 %
-Oxidant (O2)= 1.813 %
Nominal Consumption:
-Fuel = 15.22 slpm
-Air = 36.22 slpm
Exchange current [i0] = 0.027318 A
Exchange coefficient [alpha] = 0.308

Fuel cell signal variation parameters:
Fuel composition [x_H2] = 99.95 %
Oxidant composition [y_O2] = 21 %
Fuel flow rate [FuelFr] at nominal Hydrogen utilization:
-Nominal = 12.2 lpm
-Maximum = 23.46 lpm
Air flow rate [AirFr] at nominal Oxidant utilization:
-Nominal = 2400 lpm
-Maximum = 4615 lpm
System Temperature [T] = 328 Kelvin
Fuel supply pressure [Pfuel] = 1.5 bar
Air supply pressure [PAir] = 1 bar

https://www.fuelcellstore.com/hydrogen-equipment/hydrogen-storage/bl-20-metal-hydride

https://www.fuelcellstore.com/bl-60-metal-hydride

For the selection of the three hydrogen tanks, the cost of each one has entered HOMER Software inputs based on their sizes (kg) as listed in the table below.

Costs

| Size (kg) | Capital ($) | Replacement ($) | O&M ($/yr) |
|---|---|---|---|
| 0.307 | 2685 | 895 | 0 |
| 0.400 | 2745 | 915 | 0 |
| 0.636 | 3963 | 1321 | 0 |

And the capacity and pressure are listed below:

*For 0.307 Kg*

Hydrogen Capacity of 20-21 standard liters (0.76-0.80 scf)

Hydrogen Pressure when Charging or Discharging is 1-12 bar at room temperature

*For 0.400 Kg*

Hydrogen Capacity of 30-34 standard liters (1.14-1.3 scf)

Hydrogen Pressure when Charging or Discharging is 1-12 bar at room temperature

*For 0.636 Kg*

Hydrogen Capacity of 60-69 standard liters (2.28-2.64 scf)

Hydrogen Pressure when Charging or Discharging is 1-12 bar at room temperature